\documentclass[graybox]{svmult}

\usepackage{mathptmx}       
\usepackage{helvet}         
\usepackage{courier}        
\usepackage{type1cm}        
\usepackage{makeidx}         
\usepackage{graphicx}        
\usepackage{color}
\usepackage{multicol}        
\usepackage[bottom]{footmisc}

\usepackage{amsmath}
\usepackage{amssymb}
\usepackage{bm}

\makeindex             

\newcommand{\romB}{{\rm B}}

\newcommand{\romd}{{\rm d}}
\newcommand{\rome}{{\rm e}}
\newcommand{\romi}{{\rm i}}
\newcommand{\romp}{{\rm p}}

\newcommand{\VECr}{{\bm{r}}}
\newcommand{\VECq}{{\bm{q}}}

\newcommand{\kappab}{\overline{\kappa}}

\newcommand{\etal}{\emph{et}$\;$\emph{al}$.$}
\newcommand{\eg}{\emph{e}$.\,$\emph{g}$.$}
\newcommand{\ie}{\emph{i}$.\,$\emph{e}$.$}

\newcommand{\gframe}{\Gamma_{\mbox{\tiny frame}}}
\newcommand{\gfluc}{\Gamma_{\mbox{\tiny fluc}}}
\newcommand{\gbare}{\Gamma_{\mbox{\tiny 0}}}

\newcommand{\FS}[1]{\textcolor{black}{#1}}
\newcommand{\MD}[1]{\textcolor{black}{#1}}
\newcommand{\CP}[1]{\textcolor{black}{#1}}

\begin{document}

\title*{Computational studies of biomembrane systems: Theoretical considerations, simulation models, and applications}\titlerunning{Computational studies of biomembrane systems}

\author{Markus Deserno, Kurt Kremer, Harald Paulsen, Christine Peter, Friederike Schmid}

\institute{\MD{Markus Deserno}
\at Department of Physics, Carnegie Mellon University, 5000 Forbes Ave., Pittsburgh, PA 15213, USA, \email{deserno@andrew.cmu.edu}
\and
Kurt Kremer \at Max-Planck-Institute for Polymer Research, Ackermannweg 10, 55128 Mainz, Germany,
\email{kremer@mpip-mainz.mpg.de}
\and
Harald Paulsen \at Department of Biology, University of Mainz, Johannes-von-M\"uller-Weg 6, 55128 Mainz, Germany,
\email{paulsen@uni-mainz.de}
\and
\CP{Christine Peter} \at Department of Chemisty, University of Konstanz,  Universit\"atsstra\ss e 10, 78457 Konstanz, Germany,
\email{christine.peter@uni-konstanz.de}
\and
\FS{Friederike Schmid} \at Institute of Physics, University of Mainz, Staudingerweg 9, 55128 Mainz, Germany,
\email{friederike.schmid@uni-mainz.de}
}

%
%

\maketitle


\abstract{This chapter summarizes several approaches combining theory, simulation and experiment that aim for a better understanding of phenomena in lipid bilayers and membrane protein systems, covering topics such as lipid rafts, membrane mediated interactions, attraction between transmembrane proteins, and aggregation in biomembranes leading to large superstructures such as the light harvesting complex of green plants.
After a general overview of theoretical considerations and continuum theory of lipid membranes we introduce different options for simulations of biomembrane systems, addressing questions such as: What can be learned from generic models? When is it expedient to go beyond them? And what are the merits and challenges for systematic coarse graining and quasi-atomistic coarse grained models that ensure a certain chemical specificity?
}

\section{Introduction}\label{sec:1}

Lipid bilayers and membrane proteins  are one important class of biological systems where the relationship between single molecule properties and the behavior of complex  nanoscopically structured materials has been under intense investigation  for a long time. In the present review we address how approaches combining theory, simulation and experiment may help us gain a better understanding of phenomena in biomembranes.
A general overview of theoretical considerations and continuum theory of lipid membranes is given and different modeling and simulation approaches to biomembrane systems are introduced.
In particular we introduce several generic lipid simulation models and show, how these models can help us understand material properties of lipid bilayers such as bending and Gaussian curvature modulus, or membrane tension,
discuss timely topics such as lipid rafts, membrane-protein interactions, and curvature mediated interactions between proteins. These fundamental theoretical and modeling investigations are important to understand the principles that govern aggregation phenomena in biological membranes that lead to large superstructures such as the light harvesting complex of green plants. In the last section of this chapter we give an overview of multiscale modeling approaches that try to go beyond generic lipid and protein models and attempt at ensuring a certain chemical specificity while still benefiting from the time- and length-scale advantages of coarse grained simulations. Finally we conclude with the example of the light harvesting complex of green plants, for which we show first steps toward a multiscale simulation model that allows to go back and forth between a coarse grained and an atomistic level of resolution and therefore permits immediate comparison to atomic level experimental data.

\section{Theory and simulation of lipid bilayers}\label{sec:2}

To provide a basis for both the theoretical ideas and the computational
techniques which we will discuss in this chapter, we  start by reminding
the reader of some essential concepts. Sec.~\ref{ssec:continuum-elasticity}
 reviews some basic aspects of the Helfrich Hamiltonian.
Sec.~\ref{ssec:CG-models}
introduces three coarse-grained membrane models that will be used in the
remainder of this chapter.
In Secs.\ \ref{ssec:material-parameters} and \ref{ssec:surface-tension}, 
we  discuss the bending moduli and the surface tension of membranes
in more detail, and finally comment on multicomponent membranes in
Sec.\ \ref{ssec:rafts}.

\subsection{Basic concepts}

\subsubsection{Continuum elasticity of lipid membranes}
\label{ssec:continuum-elasticity}

Lipid molecules are amphipathic: they consist of a hydrophilic head group and
typically two hydrophobic (fatty acid) tails. Yet, despite their amphipathic
nature, lipid molecules dissolved in water have an extremely low critical
aggregate concentration (nanomolar or even smaller \cite{SmithTanford72}),
and thus under most common conditions lipids spontaneously aggregate. Since
the roughly cylindrical shape of lipids leads to two-dimensional self
assembly, thermodynamic considerations \cite{IMN76} show that---in contrast
to the finite size of spherical and wormlike micelles---a single macroscopic
aggregate containing almost all of the lipids will form: a two-dimensional
bilayer membrane. Its lateral dimensions can exceed its thickness by several
orders of magnitude.

\subsubsection{The Helfrich Hamiltonian}

If lipid membranes are subjected to lateral tension, they typically rupture
at stresses of several mN/m, with a remarkably low rupture \emph{strain} of
only a few percent \cite{Rawicz-etal-00}. At large sca\-les and moderate
tensions it is hence an excellent approximation to consider membranes as
largely unstretchable two-dimensional surfaces. Their dominant soft modes are
not associated with stretching but with \emph{bending} \cite{Canham70,
Helfrich73, Evans74}. Within the well-established mathematical framework
developed by Helfrich \cite{Helfrich73}, the energy of a membrane patch
$\mathcal{P}$, amended by a contribution due to its boundary $\partial
\mathcal{P}$ \cite{Helfrich74}, is expressible as
\begin{equation}
E[\mathcal{P}] = \int_{\mathcal{P}}\romd A\;\left\{\frac{1}{2}\kappa (K-K_0)^2 + \kappab K_{\rm G} \right\} 
+ \oint_{\partial\mathcal{P}} \gamma \ . \label{eq:Helfrich-Hamiltonian}
\end{equation}
Here, $K=c_1+c_2$ and $K_{\rm G}=c_1\,c_2$ are the total and Gaussian
curvature, respectively, and the $c_i$ are the local principal curvatures of
the surface \cite{Kreyszig, doCarmo}. The inverse length $K_0$ is the
spontaneous bilayer curvature, showing that the first term quadratically
penalizes the deviation between total and spontaneous
curvature.\footnote{Observe that $1/K_0$ is \emph{not} the optimal radius
$R_{\rm opt}$ of a spherical vesicle. Minimizing the energy per area with
respect to $K$ shows that instead this radius is given by $R_{\rm opt} K_0 =
2+\kappab/\kappa$.} The parameters $\kappa$ and $\kappab$ are the bending
modulus and Gaussian curvature modulus, respectively, and they quantify the
energy penalty due to bending. Finally, the parameter $\gamma$ is the free
energy of an open membrane edge and thus referred to as the edge tension.

\subsubsection{Refining the Helfrich model}
\label{ssec:helfrich-refined}

While the Helfrich Hamiltonian provides a successful framework for describing
the large-scale structure and geometry of fluid membranes, it is not designed
for modeling membranes on smaller length scales, i.e., of the order of the
membrane thickness. Several more refined continuum models have been proposed to
amend this situation.  Evidently, continuum descriptions are no longer
applicable at the {\AA}ngstr\"om scale. However, they still turn out to be
quite useful on length scales down to a few nanometers.

As one refinement, Lipowsky and coworkers have proposed to introduce a
separate, independent ``protrusion'' field that accounts for short wavelength
fluctuations \cite{LG93,LG93b,GGL99}.  According to recent atomistic and
coarse-grained simulations by Brandt \etal, these protrusions seem to
correspond to lipid density fluctuations within the membrane \cite{BE10,BBS11}.
Lindahl and Edholm pioneered another important refinement, which is to consider
the height and thickness variations of membranes separately \cite{Lindahl00}.
Continuum models for membranes with spatially varying thickness have a
long-standing tradition in theories for membrane-mediated interactions between
inclusions \cite{O78,O79,H86,SM91,DPS93,DBP94,ABD96,H95,NGA98,HHW99b,
NA00,JM04} (see also Sec.~\ref{ssec:hydrophobic_mismatch}), and they can be coupled to
Helfrich models for height fluctuations in a relatively straightforward manner
\cite{BB06,BB07,WBS09}. In addition, one can include other internal degrees of
freedom, such as local tilt \cite{Fournier98,Fournier99,BKM03,MNK07,WPW11,WPW13}, as well as
membrane tension \cite{NWN10,WPW13}.

In this article, we will focus in particular on the so-called coupled monolayer
models \cite{H86,H95,NGA98,NA00,HHW99b,DPS93,DBP94,ABD96, BB06,BB07}, where
membranes are described as stacks of two sheets (monolayers), each with their
own elastic parameters.  Monolayers are bound to each other by a local harmonic
potential which accounts for the areal compressibility of lipids within the
membrane and their constant volume \cite{ABD96,BB06}.  Li \etal\ have recently
compared the elastic properties of amphiphilic bilayers with those of the
corresponding monolayers within a numerical self-consistent field study of
copolymeric membranes \cite{LiPastor2013}.  They found that the bilayer elastic
parameters can be described at an almost quantitative level by an appropriate
combination of monolayer elastic parameters. 

\subsection{Coarse-grained lipid models}\label{ssec:CG-models}

The multitude of length- and time scales that matter for biophysical membrane
processes is mirrored in a wide spectrum of computational models that have
been devised to capture these scales. These range from all-atom simulations
\cite{Scott2002, SaizBandyopadhyayKlein02, SaizKlein02, Berkowitz09,
NimelaHyvonenVattulainen09} up to dynamically triangulated surfaces
\cite{GompperKroll97, GompperKroll98, KumarGompperLipowsky01,
NoguchiGompper04} and continuum models \cite{AtzbergerKramerPeskin07,
Brown-elastic-review08}.  The region in-between is becoming increasingly
populated by a wealth of different \emph{coarse-grained} (commonly
abbreviated ``CG'') models, which capture different aspects of a very complex
physical situation, and a number of excellent reviews exists that provide a
guide to the literature \cite{VenturoliSperottoKranenburgSmit06,
MullerKatsovSchick06, Brannigan-review06, MarrinkVriesTieleman2009,
BennunHoopesXingFaller09, Deserno09, Noguchi-review-09, Schmid-review-09}.

Besides their chosen level of resolution, CG models can also be classified by
the ``spirit'' in which they approach a physical situation: If the focus lies
on generic mechanisms that are thought to be quite universal in their reach,
there is no need to construct models that faithfully relate to every aspect of
some particular lipid. Instead, one creates ``top-down'' models based on the
presumed principles underlying the generic mechanisms of interest. For
instance, if one wishes to understand how a bilayer membrane interacts with a
colloidal particle that is much bigger than the thickness of the membrane,
relevant aspects of the situation will likely include the fluid
curvature-elastic response of bilayer lipid membranes, but probably not the
hydrogen bonding abilities of a phosphatidylethanol head group. If, in
contrast, one wishes to understand how mesoscopic membrane properties emerge
from specific properties of their microscopic constituents, the aim is instead
to construct ``bottom-up'' models whose key design parameters follow in a
systematic way from those of a more finely resolved model. For instance, if one
wishes to understand how those hydrogen bonding abilities of a
phosphatidylethanol head group impact the mesoscopic phase behavior of mixed
bilayers, it will not do to simply guess a convenient head group interaction
potential, even if it is eminently plausible. The latter philosophy goes under
various names, such as ``systematic coarse-graining'' or ``multiscaling'' and
again excellent literature and resources exists that cover this field
\cite{FMP-review02, IzvekovVoth05, praprotnik05, praprotnik07, praprotnik08,
delgado-buscalioni08, MSC-I-08, MSC-II-08, MSC-III-09, delgado-buscalioni09,
PeterKremer09, poblete10, Voth-CG-book, VOTCA09, MSC-LuVoth-12}.

The top-down and bottom-up approaches are not necessarily mutually exclusive.
It is conceivable that certain aspects of the science are systematically
matched, while others are accounted for in a generic way by using intuition
from physics, chemistry, mathematics, or other pertinent background knowledge.
Conversely, this also means that what any given model can qualitatively or
quantitatively predict depends greatly on the way in which it has been
designed; there is no universally applicable CG model. Stated differently,
systematically coarse-grained models will not be accurate in every prediction
they make, and generic models can be highly quantitative and experimentally
testable. One always needs to know what went into a given model to be able to
judge the reliability of its predictions.

In the following, (\ref{ssec:Cooke} -- \ref{ssec:MARTINI}), we will review the
basics of three particular CG models that will feature in the remainder of this
paper. The choice of models is not meant to imply a quality statement but
merely reflects our own experience and work.

\subsubsection{Cooke model}\label{ssec:Cooke}

The Cooke model \cite{Deserno2005, Cooke05} is a strongly coarse-grained
top-down lipid model in which every single lipid is represented by three
linearly connected beads (one for the head group, two for the tail) and solvent
is implicitly accounted for through effective interactions. It is purely based
on pair interactions and therefore very easy to handle. Its main tuning
parameters are the temperature, and the range $w_{\rm c}$ of the effective
cohesion that drives the aggregation of the hydrophobic tail beads. One might
also change the relative size between head- and tail-beads to control the
lipids' spontaneous curvature \cite{Cooke06}. The bead size $\sigma$ serves
as the unit of length, the potential depth $\epsilon$ as the unit of energy.
For the common choice $k_\romB T/\epsilon=1.1$ and $w_{\rm c}/\sigma=1.6$
lipids spontaneously assemble into fluid membranes with an area per lipid of
about $1.2\,\sigma^2$ and a bending rigidity $\kappa\approx 12.8\,k_\romB T$
(but rigidities between $3\,k_\romB T$ and $30\,k_\romB T$ can be achieved
without difficulty), and an elastic ratio of $\kappab/\kappa\approx -0.92$
\cite{HuBriguglioDeserno12}.

\subsubsection{Lenz model}\label{ssec:Lenz}

Like the Cooke model, the Lenz model \cite{SDL07} is a generic model for
membranes, but it has been designed for studying internal phase transitions.
Therefore, it puts a slightly higher emphasis on conformational degrees of
freedom than the Cooke model. Lipids are represented by semiflexible linear
chains of seven beads (one for the head group, six for the tail), which interact with
truncated Lennard-Jones potentials. Model parameters such as the chain
stiffness are inspired by the properties of hydrocarbon tails \cite{DS01}. The
model includes an explicit solvent, which is, however, modeled such that it is
simulated very efficiently: It interacts only with lipid beads, not with itself
(``phantom solvent'' \cite{LS05}).

The model reproduces the most prominent phase transitions of
phospholipid monolayers \cite{DS01} and bilayers \cite{LS07}. In particular,
it reproduces a main transition from a fluid membrane phase ($L_{\alpha}$) 
to a tilted gel phase ($L_{\beta'}$) with an intermediate ripple phase
($P_{\beta'}$), in agreement with experiments. The elastic parameters
have been studied in the fluid phase and are in reasonable agreement
with those of saturated DPPC (dipalmitoyl-phosphatidylcholine) bilayers.
Recently, the Lenz model was supplemented with a simple cholesterol 
model \cite{MVS13}. Cholesterol molecules are taken to be shorter and 
stiffer than lipids, and they have a slight affinity to lipids.
Mixtures of lipids and cholesterol were found to develop nanoscale
raft domains \cite{MVS13}, in agreement with the so-called ``raft hypothesis'' 
\cite{Pike06}. As a generic model that reproduces nanoscale structures
in lipid membranes (ripple states and rafts), simulations of the Lenz 
model can provide insight into the physics of nanostructure formation 
in lipid bilayers. This will be discussed in more detail in 
Sec.~\ref{ssec:rafts}.

\subsubsection{MARTINI model}\label{ssec:MARTINI}

The MARTINI model for lipids \cite{Marrink04, marrink_martini_2007} is a hybrid
between a top-down and a bottom-up model: approximately four heavy atoms are
mapped to a single CG bead, and these CG beads come in a variety of types,
depending on their polarity, net charge, and the ability to form hydrogen
bonds. The systematic aspect of MARTINI largely derives from the fact that the
non-bonded interactions between these building blocks (shifted Lennard-Jones
and possibly shifted Coulomb potentials) have been parameterized to reproduced
most of the thermodynamics correctly, especially the partitioning free energy
between different environments, such as between aqueous solution and oil. Given
a particular molecule, a judicious choice of assignments from groups of heavy
atoms to MARTINI beads, together with standard bonded interactions (harmonic,
angular, and dihedral potentials) leads to the CG version of a molecule.

The complete MARTINI force field encompasses more than lipids and sterols
\cite{Marrink04, marrink_martini_2007}; it is currently also available for
proteins \cite{MARTINI-proteins08}, carbohydrates
\cite{MARTINI-carbohydrates09} and glycolipids \cite{MARTINI-glyco13}. The
far-reaching possibilities for looking at multicomponent systems without the
need to explicitly cross-parametrize new interactions have substantially
contributed to the attractiveness of this force field. Of course, care must
still be taken that one's mapping onto the CG level is overall consistent and
chemically meaningful: Even though the non-bonded interactions are derived from
a single guiding principle, which is both conceptually attractive and
computationally powerful, there is no guarantee that it will under all
circumstances work for one's particular choice of system and observable, so it
is up to the user to perform judicious sanity checks. After all, with great
power there must also come -- great responsibility \cite{spiderman}.

\subsection{Obtaining material parameters}
\label{ssec:material-parameters}

The Hamiltonian (\ref{eq:Helfrich-Hamiltonian}) is an excellent
phenomenological description of fluid membranes, but it doesn't predict the
material parameters entering it, which must instead come from experiment or
simulation. Let us briefly list a number of ways in which this is achieved,
both in experiment and in simulation.

The bending modulus $\kappa$ is measured by techniques such as monitoring the
thermal undulations of membranes \cite{Brochard75, BGP76, Schneider84a,
Schneider84b, Faucon89, Henriksen04}, probing the low-tension stress-strain
relation \cite{Evans90}, X-ray scattering \cite{Liu04, Chu05, TNagle07,
Pan08}, neutron spin echo measurements \cite{Pfeiffer93, Takeda99,
Rheinstaedter06} (note however the caveats raised by Watson and Brown
\cite{Watson10}), or pulling thin membrane tethers \cite{Bo89, Cuvelier05,
Tian08}. In simulations, monitoring undulations \cite{GGL99,
Lindahl00, Ayton02, Farago03, Marrink04, WF05, Deserno2005, Cooke05,
BB06, Wang10a, ShibaNoguchi11} or orientation fluctuations
\cite{Watson12}, measuring tensile forces in tethers \cite{Harmandaris06,
Arkhipov08, ShibaNoguchi11}, and buckling \cite{Noguchi-buckle-2011,
HuDigginsDeserno13} have been used successfully.

The Gaussian curvature modulus $\kappab$ is much harder to obtain, since by
virtue of the Gauss-Bonnet theorem \cite{Kreyszig, doCarmo} the surface
integral over the Gaussian curvature $K_{\rm G}$ depends only on the
\emph{topology} and the \emph{boundary} of the membrane patch $\mathcal{P}$.
Hence, one needs to change at least one of them to access the
Gaussian curvature modulus. It therefore tends to be measured by looking at
the transitions between topologically different membrane phases (e.g.\ the
lamellar phase L$_\alpha$ and the invented cubic phase Q$_{\rm II}$)
\cite{Siegel04, Siegel06, Siegel08, Templer98} or the shape of
phase-separated membranes in the vicinity of the contact line
\cite{Baumgart05, Semrau08} (even though the latter strictly speaking only
gives access to the \emph{difference} in Gaussian moduli between the two
phases). In Sec.~\ref{ssec:gaussian-modulus} we will briefly present a
computational method that obtains $\kappab$ from the closure probability of
finite membrane patches \cite{HuBriguglioDeserno12,
HuDeJongMarrinkDeserno12}. 

To measure the edge tension requires an open edge, and in experiments this
essentially means looking at pores \cite{Taupin75, Zhelev93, Genco93,
Karatekin03}. This also works in simulations \cite{Farago03, Marrink04,
WF05, Deserno2005, Cooke05}, but it tends to be easier to create straight
bilayer edges by spanning a ``half-membrane'' across the periodic boundary
conditions of the simulation box \cite{Briels04a, Wohlert06, Kindt04,
Wang10b}. 

The spontaneous curvature $K_0$ usually vanishes due to bilayer up-down
symmetry, but could be measured by creating regions of opposing spontaneous
curvature and monitor the curvature this imprints on the membrane
\cite{WangHuZhang}, or by measuring the shape of a spontaneously curved
membrane strip \cite{ShibaNoguchi11}. 

Since curvature elasticity is such an important characteristic of lipid
membranes, obtaining the associated moduli has always seen a lot of
attention. Let us therefore provide a few more details on some classical and
some more recent computational strategies to measure them. Shiba and Noguchi
\cite{ShibaNoguchi11} also provide a detailed recent review.

\subsubsection{Bending modulus}
\label{ssec:bending-modulus}

The shape of essentially flat membranes stretched across the periodic
boundary conditions of a simulation box can be described by specifying their
vertical displacement $h(\VECr)$ above some horizontal reference plane, say
of size $L\times L$. In this so-called Monge parametrization the bending
contribution due to the total curvature term (ignoring for now on the
spontaneous curvature $K_0$) is given by \begin{eqnarray} \int \romd
A\;\frac{1}{2}\kappa\,K^2 & = & \frac{1}{2}\kappa\int_{[0,L]^2}\romd^2 r \;
\sqrt{1+(\nabla h)^2}\;\left(\nabla\cdot\frac{\nabla h}{\sqrt{1+(\nabla
h)^2}}\right)^2 \label{eq:Monge-1} \\ & = &
\frac{1}{2}\kappa\int_{[0,L]^2}\romd^2 r \; \left\{ (h_{ii})^2
-\frac{1}{2}(h_{ii})^2h_jh_j -2 h_{ii}h_jh_{jk}h_k + \mathcal{O}(h^6)
\right\} \ , \label{eq:Monge-2} \end{eqnarray} where the indices are
short-hand for derivatives: $h_i=\partial h/\partial \VECr_i$, etc.  The
first square root expression in Eqn.~(\ref{eq:Monge-1}) is the metric
determinant that accounts for the increased area element if the surface is
tilted. The expression following it is the total curvature in Monge gauge.
Evidently the Helfrich Hamiltonian is highly nonlinear in this
parametrization! Hence, one frequently expands the integrand for small $h$,
as is done in the second line. The first term, $\frac{1}{2}\kappa
(h_{ii})^2=\frac{1}{2}\kappa(\Delta h)^2$ is quadratic in $h$ and thus gives
rise to a harmonic theory, which is referred to as ``linearized Monge
gauge''. The majority of all membrane work relies on this simplified version.
However, the higher order terms occasionally matter: They are for instance
responsible for the renormalization of the bending rigidity by thermal shape
undulations \cite{Helfrich85, PelitiLeibler85, Foerster86, Kleinert1986}.

Upon Fourier-transforming $h(\VECr)=\sum_{\VECq}
\tilde{h}_\VECq\,\rome^{\romi\VECq\cdot\VECr}$ and restricting the functional
to quadratic order we obtain the transformed Hamiltonian $E[\tilde{h}_\VECq]
= L^2\sum_\VECq \frac{1}{2}\kappa q^4 |\tilde{h}_\VECq|^2$, which shows that
the modes $\tilde{h}_\VECq$ are independent harmonic oscillators. The
equipartition theorem then implies that $\langle |\tilde{h}_\VECq|^2\rangle =
k_\romB T/L^2\kappa q^4$, and thus fitting to the spectrum of thermal
undulations gives access to $\kappa$. Unfortunately, there are several
difficulties with this picture (see, e.g., the recent review
\cite{Schmid13}).  The simple expression can only be expected to hold for
sufficiently small wave vectors, since at small length scales local bilayer
structure will begin to matter. For instance, it is well known that lipid
tilt fluctuations contaminate the undulation spectrum \cite{MNK07,
WPW11}.  The situation becomes ever more complicated in low
temperature phases that exhibit hexatic order \cite{NP87,DR92} or permanent
tilt \cite{PL95,P96}. In such cases, the fluctuation spectrum shows no sign
of a  $\langle |\tilde{h}_\VECq|^2\rangle \propto 1/ q^4$ behavior up to
length scales of at least 40 nanometers \cite{WestSchmid10}. The most
obvious way out is to simulate larger systems and thus gain access to smaller
wave vectors, but unfortunately these modes decay exceedingly slowly. For
overdamped Brownian dynamics with a friction constant $\zeta=L^2\zeta_0$ one
finds $\zeta\dot{\tilde{h}}_\VECq = -\partial
E[\tilde{h}_\VECq]/\partial\tilde{h}_\VECq = -L^2\kappa
q^4\,\tilde{h}_\VECq$, showing that modes exponentially relax with a time
constant $\tau=\zeta_0/\kappa q^4$ that grows quartically with the wave
length.  Accounting for hydrodynamics turns this into a cubic dependence,
$\tau= 4\eta/\kappa q^3$ \cite{Kramer71,Brochard75,
SeifertLanger93,ZilmanGranek96}, where $\eta$ is the solvent viscosity, but
the situation is still uncomfortable: when Lindahl and Edholm
\cite{Lindahl00} simulated 1024 DPPC lipids in a $20\,{\rm nm}$ square
bilayer, their measured value $\kappa=4\times 10^{-20}\,{\rm J}$ implies
$\tau\simeq 3.2\,{\rm ns}$ for the slowest (and most informative) mode, not
much smaller than the overall $10\,{\rm ns}$ total simulation time.

While measuring $\kappa$ from the undulation spectrum is possible, there is a
more basic concern with such an approach: one tries to  measure a modulus
with a value typically around $20\,k_\romB T$ by using thermal fluctuations
of order $k_\romB T$ to excited the bending modes, which of course makes it
quite challenging to get a signal to begin with.\footnote{It is easy to see
that $\delta h \equiv \langle h(\VECr)^2\rangle^{1/2} = L\sqrt{k_\romB
T/16\pi^3\kappa}\approx L/100$ (assuming $\kappa\simeq 20\,k_\romB T$), which
is a few {\AA}ngstr\"om for typical simulation sizes.} An obvious alternative is
to \emph{actively} bend membranes and directly measure their curvature
elastic response. There are clearly many ways to deform a membrane; here we
will describe two possibilities which have been proposed in the past as
convenient methods for obtaining the bending modulus.

Harmandaris and Deserno \cite{Harmandaris06} proposed a method that relies on
simulating cylindrical membranes. Imagine a membrane of area $A$ that is
curved into a cylinder of curvature radius $R$. Its length $L$ satisfies
$2\pi R L=A$, and the curvature energy per area of this membrane is
\begin{equation}
e  = \frac{1}{2}\kappa \frac{1}{R^2} = \frac{1}{2}\kappa\left(\frac{2\pi L}{A}\right)^2 \ .
\end{equation}
Since
changing the length of the cylinder at constant area will also change the
curvature radius, and thus the bending energy, there must be an axial force
$F$ associated with this geometry. Its value is
\begin{equation}
F = \left(\frac{\partial eA}{\partial L}\right)_A = A\;\kappa\,\frac{2\pi L}{A}\;\frac{2\pi}{A} = \frac{2\pi\kappa}{R} \ . \label{eq:tether-force}
\end{equation}
Hence, measuring both the axial force and the cylinder radius yields the bending
modulus as $\kappa=FR/2\pi$. Notice that within quadratic curvature
elasticity the radius of the cylinder does not matter: Both small and large
radii will lead to the same modulus. In other words, $FR$ is predicted to be
a constant. Of course, it is conceivable that higher order corrections to the
Helfrich Hamiltonian (\ref{eq:Helfrich-Hamiltonian}) matter once curvatures
become really strong. For the present geometry there is only one term, which
enters at quartic order, and one would write a modified energy density
$e=\frac{1}{2}\kappa K^2+\frac{1}{4}\kappa_4K^4$. This modified functional
leads to $FR/2\pi=\kappa+\kappa_4/R^2\equiv \kappa_{\rm eff}(R)$, which can
be interpreted as an effective curvature dependent bending modulus.
Simulations using different models with different levels of resolution have
indeed both seen a small dependence of $\kappa_{\rm eff}$ on $R$
\cite{Harmandaris06, ShibaNoguchi11}. They find \emph{softening} at high curvature, which would indicate that $\kappa_4$ is
negative.  In contrast, Li \etal\ \cite{LiPastor2013} recently studied
the elastic properties of self-assembled copolymeric bilayers by
self-consistent field theory in cylindrical and spherical geometry, and found
$\kappa_4$ to be positive. The details of nonlinear elastic corrections
thus depend on specifics of the model under study, but the present
studies suggest that as long as the radius of curvature is bigger than a few
times the membrane thickness, these corrections are negligible. For
example, Li \etal\ \cite{LiPastor2013} found the
deviations from linear to be less than 2\% both in the cylinder and sphere
geometry, as long as the reduced curvature was less than $K_0 d = 0.6$ (where $d$
is the bilayer thickness).

The cylinder stretching protocol appears to work very well for simple
solvent-free membrane models \cite{Harmandaris06, Arkhipov08,
ShibaNoguchi11}, but with more refined models this method suffers from two
drawbacks, both related to the equilibration of a chemical potential. First,
the cylinder separates the simulation volume into an ``inside'' and an
``outside''. If solvent is present, its chemical potential must be the same
in these two regions, but for more highly resolved models the solvent
permeability through the bilayer is usually too low to ensure automatic
relaxation. Second, the chemical potential of lipids also has to be the same
in the two bilayer leaflets, and again for more refined models the lipid
flip-flop rate tends to be too low for this to happen spontaneously.

To circumvent this difficulty, Noguchi has recently proposed to instead
simulate a buckled membrane as an example of an actively imposed deformation
\cite{Noguchi-buckle-2011}. This solves both problems simultaneously: Neither
does the buckle divide the simulation box into two distinct
compartments,\footnote{Observe that the part of the membrane above the buckle
and the part below the buckle can be connected through the periodic boundary
of the simulation box.} nor is lipid equilibration across leaflets a big
concern, since for symmetry reasons both leaflets  are identical (at least
for a ``ground state buckle'') and thus ensuring that the same number of
lipids is present in both leaflets is a good proxy. The theoretical analysis
of the expected forces is a bit more complicated compared to the cylinder
setup, but it can be worked out exactly even for buckles deviating strongly
from ``nearly flat''.  Hu et al.\ have recently provided systematic series
expansions for the buckling forces in terms of the buckling strain. If a
membrane has originally a length $L$ and is buckled to a shorter length
$L_x$, then the force $f_x$ per length along that membrane as a function of
strain $\gamma=(L-L_x)/L$ can be written as \cite{HuDigginsDeserno13}
\begin{equation}
f_x = \kappa \left(\frac{2\pi}{L}\right)^2\left[1+\frac{1}{2}\gamma+\frac{9}{32}\gamma^2+\frac{21}{128}\gamma^3+\frac{795}{8192}\gamma^4 + \frac{945}{16384}\gamma^5 + \cdots\right] \ .
\end{equation}
Notice that the force does not vanish for $\gamma\rightarrow 0$, which is the
hallmark of a buckling transition. Hu et al.\ \cite{HuDigginsDeserno13} also
estimate the fluctuation correction on this result and find it to be very
small; they apply this method to four different membrane models, ranging from
strongly coarse grained to essentially atomistic, and argue that it is
reliable and efficient.

\begin{figure}[t]
\sidecaption[t]
\includegraphics[scale=.12]{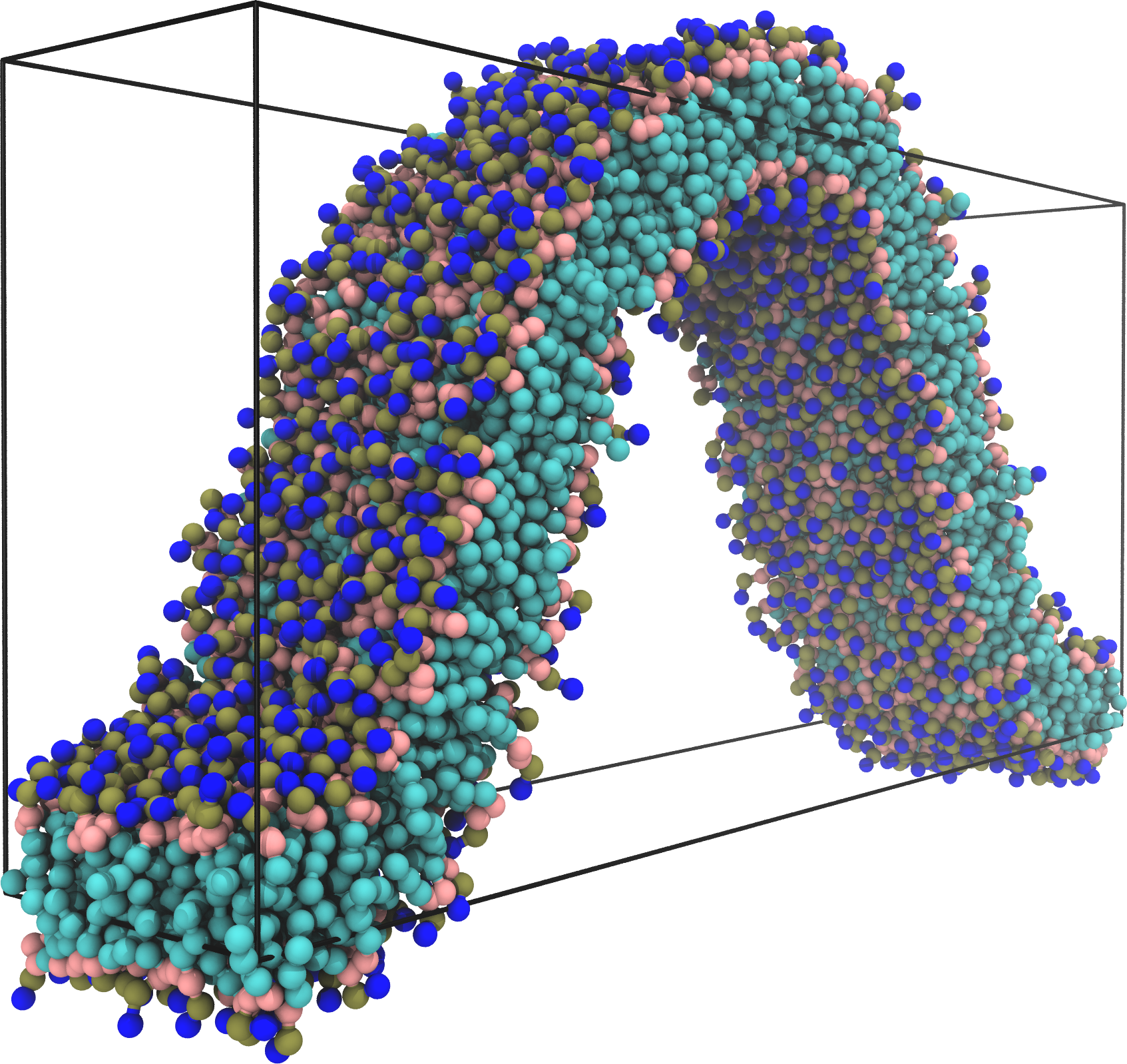}
\caption{Illustration of a buckling simulation using the MARTINI model for
DMPC (which uses 10 beads per lipid) \cite{HuDigginsDeserno13}. This
particular membrane consists of 1120 lipids and is compressed at a strain
$\gamma=0.3$, which gives it an amplitude of approximately 22\%\ of the box
length. To suppress membrane deformations in the second direction, the width
of the box is chosen much smaller than its length. }\label{fig:buckle}
\end{figure}

\vspace*{1em}

\subsubsection{Gaussian modulus}
\label{ssec:gaussian-modulus}

To measure the Gaussian curvature modulus $\kappab$ directly, the
Gauss-Bonnet theorem \cite{Kreyszig, doCarmo} forces one to either change the
topology or the boundary of a membrane patch. Recently Hu et al.\
\cite{HuBriguglioDeserno12, HuDeJongMarrinkDeserno12} suggested a way to
achieve this. Consider a circular membrane patch of area $A$. Being flat, its
energy stems from the open edge at its circumference. The patch could close
up into a vesicle in order to eliminate the open edge, but now it carries
bending energy.  If we imagine that transition proceeding through a sequence
of conformations, each one resembling a spherical cap of curvature $c$, then
the excess energy of such a curved patch (compared to the flat state) is
given by \cite{Helfrich74, Fromherz83}
\begin{equation}
  \frac{\Delta E(x,\xi)}{4\pi(2\kappa+\kappab)} = \Delta\tilde{E}(x,\xi) = x+\xi\left[\sqrt{1-x}-1\right] \ .
  \label{eq:energy-barrier}
\end{equation}
$\Delta E$ is scaled by the bending energy of a sphere and we defined
\begin{equation}
x=(Rc)^2
\quad,\quad
\xi = \frac{\gamma R}{2\kappa+\kappab}
\;\;\;,\quad{\rm and}\quad
R=\sqrt{\frac{A}{4\pi}} \ .
\label{eq:definitions}
\end{equation}
For $\xi>1$ the spherical state ($x=1$) has a lower energy than the flat
state ($x=0$). If $x$ is viewed as a reaction coordinate,
Eqn.~(\ref{eq:energy-barrier}) describes a nucleation process, since for
$\xi<2$ the transition from the flat to the spherical state proceeds through
an energy barrier of height $\Delta \tilde{E}^\ast=(1-\xi/2)^2$ at
$x^\ast=1-(\xi/2)^2$. Eqn.~(\ref{eq:energy-barrier}) shows that the
functional form of the nucleation barrier depends on parameters such as
moduli and system sizes only though the combination $\xi$, and all parameters
entering $\xi$ -- with the exception of $\kappab$ -- can be determined ahead
of time by other means. Hence, measuring $\kappab$ amounts to measuring the
nucleation barrier (or at least the location of the maximum). Hu et al.\
\cite{HuBriguglioDeserno12, HuDeJongMarrinkDeserno12} do this by a dynamical
process: Equilibrated but pre-curved membrane patches with some initial value
for $x$ may either flatten out or close up, depending on where on the barrier
they start. The probability for either outcome can be computed if $\Delta E$
is known \cite{vanKampen} and so $\kappab$ ends up being found through a
series of patch-closure experiments.

The results of such simulations show that $\kappab/\kappa$ is close to $-1$,
both for the Cooke model and for MARTINI DMPC (see Sec.~\ref{ssec:CG-models}
for a further discussion of these models).\footnote{The requirement that the
Hamiltonian (\ref{eq:Helfrich-Hamiltonian}) is bounded below demands
$-2\kappa \le \kappab \le 0$.}
This is compatible with experiments
\cite{Siegel04, Siegel06, Siegel08, Templer98, Baumgart05, Semrau08} but
disagrees with the only other method that has been suggested for getting the
Gaussian modulus. As first pointed out by Helfrich \cite{Helfrich81}, quite
general considerations suggest that the second moment of a membrane's lateral
stress profile is also equal to the Gaussian modulus \cite{Helfrich81,
Helfrich94, Szleifer90}:
\begin{equation}
\kappab = \int\romd z\; z^2 \Sigma(z) \ , \label{eq:kappab-stress-profile}
\end{equation}
where $\Sigma(z) = \Pi_{zz}-\frac{1}{2}[\Pi_{xx}(z)+\Pi_{yy}(z)]$ is the
position-resolved lateral stress through a membrane, whose integral is simply
the surface tension \cite{RW}. However, when applied to the Cooke model, Hu
et al.\ find $\kappab/\kappa\approx -1.7$ \cite{HuBriguglioDeserno12}, quite
a bit more on the negative side, while applying it to MARTINI DMPC (at
$30\,{\rm K}$) yields $\kappab/\kappa\approx -0.05$, much closer to zero;
MARTINI DPPC and DOPC even lead to positive Gaussian moduli. At present it is
quite unclear where this discrepancy originates from, but given that the
values obtained from the patch-closure protocol are physically more plausible
it seems likely that there is a problem with the stress approach. The latter
suspicion is also supported by the fact that a more refined theory of
bilayer elasticity \cite{GompperKlein92} predicts corrections to the right
hand side of Eqn.~(\ref{eq:kappab-stress-profile}) that depend on moments of
order parameter distributions.

\subsection{The tension of lipid membranes}\label{ssec:surface-tension}

The Helfrich Hamiltonian, Eq.\ (\ref{eq:Helfrich-Hamiltonian}), does not
include a surface tension contribution. Free membrane patches can relax and
adjust their area such that they are stress-free. In many situations, however
membranes do experience mechanical stress. For example, an osmotic pressure
difference between the inside and the outside of a lipid vesicle generates
stress in the vesicle membrane. Stress also occurs in supported bilayer
systems, or in model membranes patched to a frame. In contrast to other
quantities discussed earlier (bending stiffness etc.), and also in contrast
to the surface tension of demixed fluid phases, membrane stress is not a
material parameter. Rather, it is akin to a (mechanical or thermodynamic)
control parameter, which can be imposed through boundary conditions.

The discussion of membrane tension is complicated by the fact that there
exist several different quantities which have been called ``tension'' or
``tension like''.  For the sake of simplicity, we will restrict ourselves to
quasi-planar (fluctuating) membranes in the following. A thoughtful analysis
of the vesicle case has recently been carried out by Diamant
\cite{Diamant11}.

The first tension-like quantity in planar membranes is the lateral 
mechanical stress in the membrane discussed above. If the stress is
imposed by a boundary condition, such as, for instance, a constraint 
on the lateral (projected) area of the membrane, it is an internal 
property of the membrane system which depends, among other, on the
areal compressibility \cite{WPW13} and the curvature elasticity 
\cite{CG-02, CG-04, FP04, Guven-aux-04, MDG-05a, MDG-05b, 
MullerDeserno07, Deserno-etal07}. Alternatively, mechanical 
stress can be imposed externally. In that case, the projected area 
fluctuates, and the appropriate thermodynamic potential can be
introduced into the Helfrich Hamiltonian, 
Eq.\ (\ref{eq:Helfrich-Hamiltonian}), in a straightforward manner:  
\begin{equation}
G = E - \gframe A_\romp = 
\int_{\mathcal{P}}\romd A\;\left\{\frac{1}{2}\kappa (K-K_0)^2 
+ \kappab K_{\rm G} - \gframe \frac{\romd A_\romp}{\romd A}
\right\}.
\label{eq:helfrich_frame}
\end{equation}
Here $\gframe$ is the stress or ``frame tension'', $A_\romp$ is the projected
area in the plane of applied stress, and we have omitted the membrane edge
term. Let us consider a membrane with fixed total area $A$. Since in Monge
representation, one has $\romd A_\romp/\romd A = 1/\sqrt{1+(\nabla h)^2} \approx
1 - (\nabla h)^2/2 + \mathcal{O}(h^4) $, the last term in Eq.\
(\ref{eq:helfrich_frame}) takes the form \cite{NWN10,Schmid11}
\begin{equation}
\mbox{const} +
\frac{1}{2}\gframe \int_{A_\romp}\romd^2 r \; (\nabla h)^2  + \mathcal{O}(h^4)
\label{eq:Monge-3}
\end{equation}
(with $\mbox{const} = - \gframe A$). This is formally similar to a
surface tension term in an effective interface Hamiltonian for liquid-liquid
interfaces. The main difference is that the base $A_\romp$ of the integral
fluctuates. However, replacing this by a fixed base $\langle A_\romp \rangle$ 
only introduces errors of order $\mathcal{O}(h^4)$ \cite{Schmid11}.

From Eq.~(\ref{eq:Monge-3}), it is clear that mechanical stress
influences the fluctuation spectrum of membranes, and in particular,
one expects a $q^2$ contribution to the undulation spectrum,
$\langle |\tilde{h}_\VECq|^2\rangle^{-1} \sim  \gfluc q^2  + \kappa q^4 
+ \cdots$. This introduces the second tension-like parameter in planar
fluctuating membranes,  the ``fluctuation tension'' $\gfluc$.
According to Eq.\ (\ref{eq:Monge-3}), $\gfluc$ is identical 
to $\gframe$ up to order $\mathcal{O}(h^2)$.

Finally, the third tension-like parameter in membranes has been
introduced by Deuling and Helfrich already in 1976 \cite{Deuling76}, 
and it couples to the total area of the membrane 
\begin{equation}
E = \int_{\mathcal{P}}\romd A\;\left\{\frac{1}{2}\kappa (K-K_0)^2 
+ \kappab K_{\rm G} + \gbare \right\}.
\label{eq:helfrich_deuling}
\end{equation}
In membranes with fixed lipid area, but variable number of lipids, the ``bare
tension'' $\gbare$ is simply proportional the lipid chemical potential. For
membranes with fixed number of lipids and variable lipid area, the physical
meaning of $\gbare$ is less clear, but it can still be defined as a field
that is conjugate to $A$ in a Lagrange multiplier sense. This term also
gives rise to a $q^2$ term in the undulation spectrum, with the fluctuation
tension $\gfluc = \gbare + \mathcal{O}(h^2)$ \cite{WPW13}.

At leading (quadratic) order in $h$, the three tension-like quantities,
$\gframe$, $\gfluc$, and $\gbare$, thus have identical values. Nevertheless,
they might differ from each other due to nonlinear corrections
\cite{BGP76,DL91,CLN94,Jaehnig96,Marsh97}. For instance,
the bare tension $\gbare$ is expected to deviate from the frame tension
$\gframe$ due to the effect of fluctuations.  The exact value of the
correction depends on the ensemble and differs for systems with a fluctuating
number of lipids (variable number of undulating modes) or a fixed number of
lipids (fixed number of modes). The former case was analyzed by Cai \etal\
\cite{CLN94}, the latter case by Farago and Pincus \cite{FP03} and 
subsequently by a number of other authors \cite{Imparato06,Stecki08,FB08}. 
Interestingly, the correction has an additive component in both cases. 
Hence a stress-free membrane has a finite bare tension.

Whereas the bare tension, $\gbare$, is mostly of academic interest,
the fluctuation tension, $\gfluc$, describes actual membrane conformations.
The relation between $\gfluc$ and $\gframe$ has been discussed somewhat
controversially in the past \cite{CLN94,FP04,Imparato06,Stecki08,FB08,
Schmid11, Fournier12, Schmid12, Farago11, Diamant11}.
Cai et al \cite{CLN94} and Farago and Pincus \cite{FP04} have presented 
a very general argument why $\gframe$ and $\gfluc$ should be equal. Cai et 
al \cite{CLN94} examined the fluctuations of planar membranes with 
variable number of lipids and fixed lipid area, and proved 
$\gfluc = \gframe$ in the thermodynamic limit, if the membrane is 
``gauge invariant'', i.e., invariant with respect to a rotation
of the ``projected plane''. Farago and Pincus \cite{FP04,Farago11} developed
a similar theory for membranes with fixed number of lipids at fixed
projected area. Unfortunately, these arguments -- albeit appealing -- are 
not entirely conclusive, since the underlying assumptions can be 
questioned: The thermodynamic limit does not exist for stress-free 
planar membranes, since they bend around on length scales larger than 
the persistence length \cite{BGP76}. In the presence of stress, it does not exist 
either, strictly speaking, because the true equilibrium state is one 
where the membrane has ruptured. Furthermore, high stresses break
gauge invariance.  
Contradicting Cai \etal\ and Farago and Pincus \cite{CLN94,FP04}, a number of 
authors have claimed $\gfluc=\gbare$, \cite{Imparato06,Stecki08,FB08}
based on analytical arguments which however also relied on the existence 
of the thermodynamic limit and on other uncontrolled approximations 
\cite{Schmid11,Fournier12,Schmid12}.

Thus the relation between $\gframe$ and $\gfluc$ remains an open
question, and simulations can point at the most likely answer. For example,
if $\gfluc=\gframe$, the fluctuation tension should vanish for stress-free
membranes, i.e., the undulation spectrum should then be dominated by a
$q^4$-behavior. With a few exceptions \cite{Imparato06,Stecki08}, 
this has indeed been observed in coarse-grained or atomistic simulations 
of stress-free lipid bilayers \cite{GGL99,Lindahl00,MM01,WF05,BB06,WBS09} or 
bilayer stacks \cite{LMK03}. This would seem to rule out the alternative
hypothesis, $\gfluc=\gbare$. However, it should be noted that the undulation 
spectra have relatively large error bars and a complex behavior at higher 
$q$, as discussed in section \ref{ssec:bending-modulus}. Therefore, the
results also depend to some extent on the fit.

To overcome these limitations, accurate simulations of elastic infinitely
thin sheets with no molecular detail are useful. Recently, a 
number of such simulations have been carried out in two spatial dimensions 
(i.e., one dimensional membranes) \cite{FB08,Schmid11,Farago11}.
The results are found to depend on the ensemble. Fournier and Barbetta
studied a membrane made of hypothetical ``lipids'' with freely fluctuating
areas, only controlled by a Lagrange parameter $\gbare$. They found that
the fluctuation tension $\gfluc$ displays a complex behavior and neither 
agrees with $\gbare$ nor with $\gframe$. Schmid \cite{Schmid11} has 
considered an arguably more realistic situation where ``lipids'' have
fixed area, and either a fixed frame tension is applied, or the ``projected
area'' is kept fixed. These simulations reproduce the difference between 
$\gbare$ and $\gframe$ and indicate with high accuracy that the 
fluctuation tension is given by $\gfluc = \gframe$.
Farago \cite{Farago11} confirmed these findings in simulations at fixed 
projected area. Furthermore, he carried out reference simulations of a 
hypothetical membrane model which lacks gauge invariance, and found that
in this case, the fluctuation tension deviates from the frame tension. 
These studies support the validity of the picture originally
put forward by Cai \etal\ \cite{CLN94}: For rotationally invariant
membranes with fixed area per lipid, the fluctuation tension is
given by the frame tension.

\subsection{Membrane heterogeneity and lipid rafts}\label{ssec:rafts}

In the late 1990 several scientists put forward the suggestion that
biomembranes might not be laterally homogeneous but instead contain nanoscopic
domains---soon called ``lipid rafts''---which differ in their lipid composition
and greatly matter for numerous membrane-associated biological processes
\cite{Ahmed-etal-97, SI97, BrownLondon98-1, BrownLondon98-2, BrownLondon00}.
This idea quickly replaced the until then prevailing fluid mosaic model
\cite{SingerNicolson72}, according to which the lipid bilayer merely
constitutes a two-dimensional passive solvent that carries membrane proteins.
It created huge excitement due to many obvious biological implications and
possibilities; at the same time it has long been discussed controversially, for
instance because it took time to converge on a universally accepted definition
of what a raft is \cite{Munro03, Pike06, Hancock06, Leslie11}.

According to the lipid raft concept, biomembranes are filled with locally phase
separated, cholesterol-rich, nanoscale ``raft'' domains, which contribute to
membrane heterogeneity and play an important role in organizing the membrane
proteins.  Two aspects of this hypothesis are well-established: (i) Biological
membranes are laterally heterogeneous, and heterogeneity is important for the
function of membrane proteins, \eg\ in signaling \cite{VSM03}. (ii)
Multicomponent lipid bilayers phase separate in certain parameter regions into
a ``liquid disordered'' (ld) and a ``liquid ordered'' (lo) phase
\cite{VeatchKeller03, VK05}.
The hypothetical ``raft state'' is not phase-separated, but rather a globally
homogeneous state filled with nanodomains of sizes between 10 and 100
nanometers.  The raft concept is supported by experimental findings, \eg\ on
the mobility of certain membrane proteins \cite{PKF00}. It has been questioned
mainly due to a lack of direct evidence. Rafts are too small to be visualized
in vivo by microscopy.  Moreover, it was not clear from a theoretical point of
view why nanoscale rafts should be stable with respect to macrophase
separation. To explain this, it was proposed that rafts might be nonequilibrium
structures \cite{TSS05}, that rafts might be stabilized by the cytoplasm
\cite{YW07} or by special line-active lipids \cite{SK04,BPS09,YBS10}.
Alternatively, it was argued that ``rafts'' might simply be a signature of
critical fluctuations in the vicinity of critical points \cite{VSK07, HVK09}.

\begin{figure}[t]
\sidecaption[t]
\includegraphics[scale=0.2]{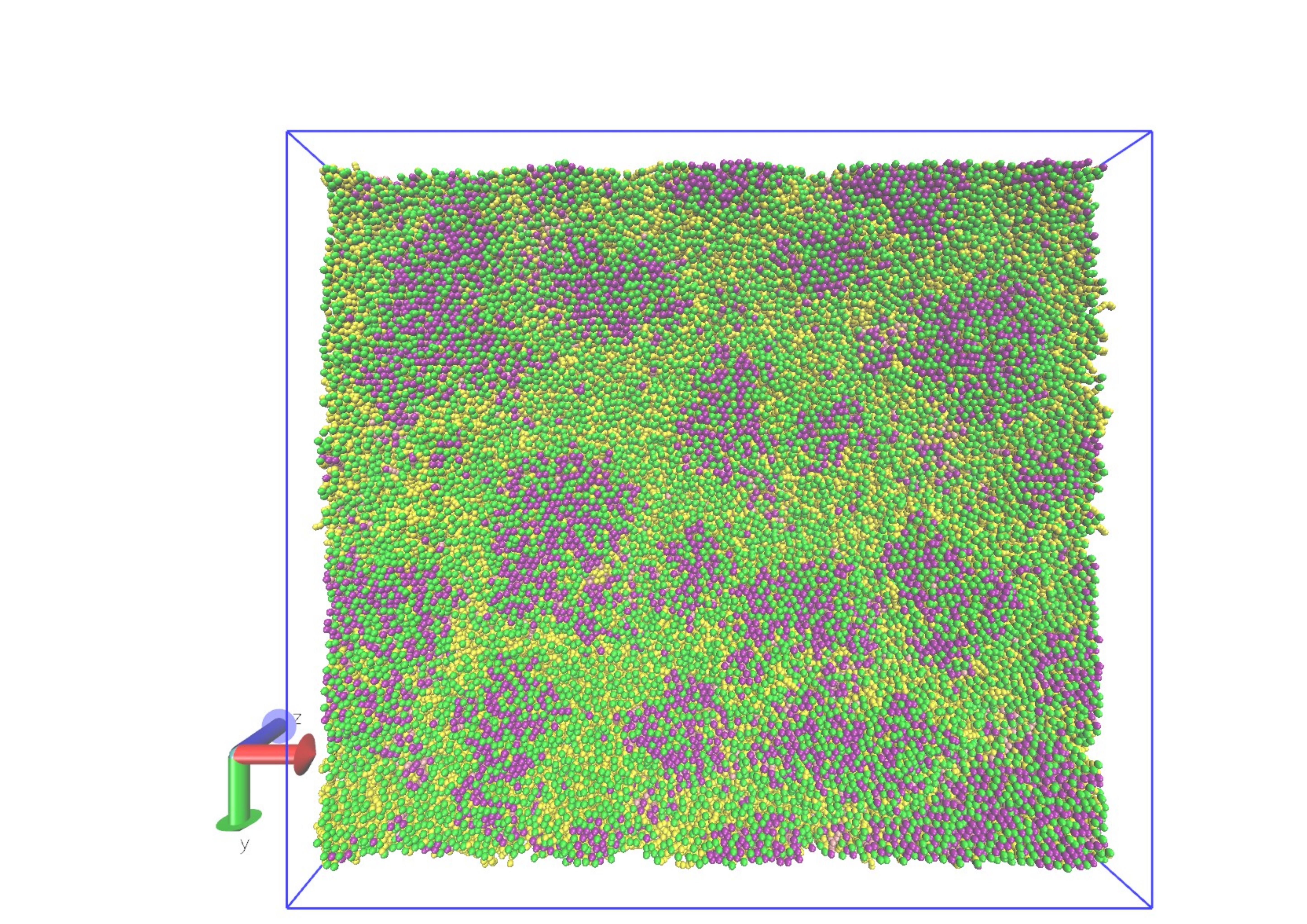}
\caption{
Rafts in a two-component lipid bilayer ($20\,000$ lipids, Lenz model).
Purple (darker) beads correspond to cholesterol, green (lighter) beads
to phospholipids. (see \protect\cite{MVS13}).
}
\label{fig:rafts}
\end{figure}

Whether a thermodynamically stable nano-struc\-tu\-red raft state could exist
in simple multicomponent membranes that do not contain special line-active
additives has remained unclear until recently.  This theoretical question mark
could be removed by recent simulations of the two-component Lenz model by
Meinhardt \etal\ \cite{MVS13}. Fig.~\ref{fig:rafts} shows a top view of a
configuration which contains microscopic cholesterol rich domains. The
simulations were carried out in a grand canonical ensemble where lipids and
cholesterol can swap identities, which excludes the possibility that the finite
domains simply reflect incomplete phase separation. The lateral structure
factor of the membranes exhibits a peak around $q \sim 0.08 \mbox{nm}^{-1}$.
Its existence shows that the clusters are not critical.  Hence, raft-like
structures can be thermodynamically stable in multicomponent membranes.  The
characteristic length scale of roughly $12\,{\rm nm}$ is compatible with the
size commonly attributed to lipid rafts in biomembranes \cite{Pike06}.

Two comments are in place here. First, it should be noted that typical ``raft
mixtures'' used for studying rafts in model membranes contain at least three
components. This is because three components seem necessary to bring about
global lateral phase separation \cite{VK05}.  Meinhardt \etal\ report raft-like
structures in a simulations of a coarse-grained model for binary mixtures, but
as in experiments, their systems do not show global phase separation between
fluid states. Likewise, there is also some experimental evidence that
nanoscopic domains may already be present in binary mixtures -- in particular
mixtures of saturated lipids (lipids with high main transition temperature) and
cholesterol. Studies based on local techniques such as ESR, NMR, or diffusitivy
measurements have indicated the existence of immiscible liquid phases
\cite{IKM87,ST91,VK05}, whereas in fluorescence microscopy, one only observes
one homogeneous phase \cite{VK05}. This suggests that these two-component
membranes phase separate on the nanoscale, while remaining homogeneous on the
global scale, and that they thus feature many of the intriguing properties
attributed to rafts.

Second, the characteristic length scale of the rafts is similar to the wave
length of the ripple state in one-component bilayers in the transition region
between the fluid and the tilted gel $L_{\beta'}$ state \cite{KTL00,SRK03}.
Experimentally \cite{KC94,KC98} and in computer simulations
\cite{KS05,VYM05,LS07,SG08,JKN10,CPM11}, modulated phases are observed in lipid
bilayers that exhibit a tilted gel state, and they are not observed in lipid
bilayers with an untilted gel state $L_\beta$ \cite{KC94,KC98,KS05,Dolezel10}.
For example, in the Lenz model, rippled states occur in the standard setup with
a mismatch between head and tail size \cite{LS07}, but they disappear if the
head size is reduced such that the tilt in the gel phase vanishes
\cite{Dolezel10}.

Meinhardt \etal\ \cite{MVS13} have proposed a joint theoretical explanation for
these findings, which is based on the coupled monolayer model (see Sec.\
\ref{ssec:helfrich-refined}). They assumed that monolayers exhibit local phase
separation into two phases with different order parameter (composition or
other), and that the spontaneous curvature of the monolayer depends on the
local order parameter. In the strong segregation limit where different phases
are separated by narrow interfaces, they showed that the line tension is
reduced in the presence of a mismatch $\Delta K_0$ between the spontaneous
curvatures of the two phases.  This is because monolayers with a spontaneous
curvature, which are forced into being planar by the apposing monolayer,
experience elastic stress, and some of that stress can be released at the
domain boundaries. The resulting negative contribution to the line tension
scales with $\kappa \: (\Delta K_0)^2$ and should be present wherever $\Delta
K_0$ is nonzero. A more detailed calculation shows that the elastic energy is
minimized for circular or stripe domains of a specific size, which is of the
order of a few nanometers. This elastic mechanism could thus stabilize rafts of
finite size for sufficiently large spontaneous curvature mismatch.

Meinhardt \etal\ also considered the weak segregation limit, where the
phase separation is incomplete, the interfaces are broad, and the free 
energy can be expanded in powers of the order parameter $\Phi$. They showed that 
the expansion has a Landau-Brazovskii form \cite{Brazovskii75},
\begin{equation}
F = \int \romd^2 r \, \left\{ \frac{g}{2}(\Delta + q_0^2)^2 \, \Phi^2
+ \frac{r}{2} \Phi^2 - \frac{\gamma}{3!} \Phi^3 + \frac{\lambda}{4!}\Phi^4\right\} \ ,
\end{equation}
with a characteristic wave vector of the order $q_0 < 1/\xi$, where $\xi$ is
the in-plane correlation length $\xi = (\kappa t_0^2/K_A)^{1/4}$ ($t_0$ is the
monolayer thickness and $K_A$ the areal compressibility).  The
Landau-Brazovskii model describes phase transitions driven by a
short-wavelength instability between a disordered and one or several ordered
phases. In mean-field approximation, it predicts a transition from a disordered
phase to one of several ordered modulated phases (lamellar or hexagonal).
Fluctuations are known to shift the order-disorder transition and to stabilize
a locally structured disordered phase {\em via} the so-called Brazovskii
mechanism \cite{Brazovskii75}. The correlation length $\xi$ sets the order of
magnitude and a lower limit for the characteristic wave length of the
structures. Inserting typical numbers for the elastic parameters of DPPC
bilayers in the fluid phase, one obtains $\xi \sim 1\,{\rm nm}$.  

The simple theory put forward by Meinhardt \etal\ accounts in a unified manner
for both ripple phases and raft states in membranes.  The prerequisites for the
formation of such modulated phases is local phase separation (e.g., in the
ripple case, between a liquid and a gel phase, or in the raft case, between a
liquid disordered and a liquid ordered phase) and curvature stress in at least
one of the two phases (typically the ordered one), resulting, e.g., from a size
mismatch between head group and tails.  In order to reproduce rippled states or
rafts, coarse-grained simulation models must meet these criteria. This is often
not the case.  For example, the standard version of the popular MARTINI model
does not have a ripple phase, because the low-temperature gel phase of
saturated phospholipids is untilted. 

\section{Membrane-protein interactions}

Biomembranes achieve their biological functions through a multitude of
mem\-bra\-ne-asso\-ci\-a\-ted proteins. Whereas the membranes were long thought
to mainly serve as a more or less inert background matrix for these proteins,
the interactions between membranes and proteins have received more and more
attention in recent years \cite{Marsh08}. Membranes can affect protein function
in several ways. The local lipid environment can immediately influence the
function of proteins---\eg, by influencing the tilt and relative position of
transmembrane domains \cite{BS12}, or by exerting local pressure on proteins
\cite{Cantor97}.  Furthermore, membranes contribute to the effective
interactions between proteins \cite{END83, BHS06, Antonny06}, and they can be
used to tune protein clustering.  In mixed membranes, the ``raft hypothesis''
mentioned in Sec.\ \ref{ssec:rafts} asserts that nanoscale lipid domains in
membranes help to organize and control protein assembly \cite{SI97, Pike06}.

Membrane-protein interactions are controlled by various factors: Local lipid
packing, local lipid concentration, membrane distortion, monolayer and bilayer
elasticity. Proteins are surrounded by a shell of lipid molecules (the lipid
annulus), which mostly interact non-specifically with the protein molecules
\cite{Lee05}.  Protein-membrane interactions are thus to a large extent
determined by the interactions of the annuli with the bulk, and
often do not depend strongly on the details of the protein sequences. 
If membrane proteins locally deform the lipid bilayer to which they are bound,
this can induce forces between them that are potentially long-ranged and quite
universal in their characteristics. The reason is that the bilayer acts as a
\emph{field} that can transmit local perturbations---and thus forces---to
distant regions. This is perfectly analogous to the way in which for instance
an electrostatic field mediates interactions between electric charges or curved
space-time mediates interactions between masses, except that a membrane seems
more tangible than the other examples. However, once we look beyond fundamental
forces towards higher level emergent phenomena, very tangible fields exist
everywhere. For instance, a rope can transduce a tensile force along its
length, and we can describe this using continuum elasticity as the underlying
``field equation''.

Just like ropes, fluid lipid membranes are continuous media at a sufficiently
coarse level of description. But their rich physical structure equips them with
several properties that can take on the role of a field, for instance:
\begin{enumerate}
\item[a)] 
\FS{
The membrane thickness can be considered as a spatially varying field
that couples to the protein content (see Sec.\ \ref{ssec:helfrich-refined}).}
\item[b)] The lipids can have a spatially varying orientation or tilt order.
\item[c)] In mixed membranes the local lipid concentrations can be viewed as a field.
\item[d)] The Hamiltonian (\ref{eq:Helfrich-Hamiltonian}) associates a characteristic energy to a given shape of a membrane, thus rendering its entire geometry a field.
\end{enumerate}

These fields differ quite substantially in their theoretical
description---con\-cen\-tra\-ti\-ons  are \emph{scalar} variables, orientations are
\emph{vectors}, and differential geometry is at heart a \emph{tensor} theory---but all
of them are known to mediate interactions. For instance, the fact that proteins
might prefer one lipid composition over another and thus aggregate
\cite{Gil-etal-97, Gil-etal-98, Reynwar08, Machta-etal-12} is central to an
important mechanism attributed to lipid rafts; and tilt-mediated protein
interactions have also been studied in multiple contexts \cite{MayBenShaul-99,
Fournier99, May-tilt-00, BKM03, Kozlovsky-etal-04, MDG-05b}. It
is even possible to describe all these phenomena within a common language
\cite{Deserno_BiomembraneFrontiers09}, using the framework of covariant surface
stresses \cite{CG-02, CG-04, Guven-aux-04, MDG-05a, MDG-05b, MullerDeserno07,
Deserno-etal07}. However, in the present review we will restrict to only two
particular example, both related to membrane elasticity: in Sec.\
\ref{ssec:hydrophobic_mismatch} we will discuss interactions due to hydrophobic
mismatch, and in Sec.\ \ref{ssec:curvature} we will look at interactions mediated by the
large-scale curvature deformation of the membrane.

\subsection{Hydrophobic mismatch}
\label{ssec:hydrophobic_mismatch}

Proteins distort or disrupt membranes, which in turn act back on proteins.
Structural perturbations contribute to protein function and are among the most
important sources of membrane-induced interactions between proteins.
Unfortunately, perturbations or transformations of lipid bilayers due to
proteins are very difficult to probe experimentally \cite{May00}.
Complementary theoretical and computer simulation studies can help to elucidate
the role of the lipid bilayer in processes such as protein aggregation and
function.

One major source of membrane-protein interactions that has been discussed
in the literature for many decades is hydrophobic mismatch
\cite{BEM91,DPS93,DBP94,ABD96,KI98,DLT99,NGA98,NA00,HHW99,PK03,BB06,BB07,
Marsh08b,CM11,SEU12}.  If the width of the hydrophobic transmembrane domain of
a protein is larger than the thickness of the lipid bilayer, the system can
respond in two ways: Either the protein tilts \cite{VSS05,PO05,HK10}, or the
membrane deforms \cite{H86,H95,NGA98}.  Both responses have biologically
relevant consequences.  On the one hand, the orientation of proteins is
believed to have a significant influence on their functionality, \eg\ in pore
formation \cite{STW09}.  Coarse-grained simulations by Benjamini and Smit have
suggested that the cross-angle distributions of packed helix complexes are
mostly determined by the tilt angle of individual helices \cite{BS12}.
Membrane deformation, on the other hand, induces effective protein-protein
interactions and provides one way to control protein aggregation
\cite{HHW99,PK03,SW10}. In experimental tilt measurements, hydrophobically
mismatched proteins were sometimes found to tilt; in other cases, the reported
tilt angles were surprisingly small compared to theoretical expectations
\cite{HB00,ORL05,VGD08}. This was partly attributed to problems with the
analysis of experimental NMR (nuclear magnetic resonance) data \cite{SEU12},
partly to the presence of anchoring residues flanking the hydrophobic
transmembrane domains, which might prevent tilting through a variety of
mechanisms \cite{ORL05,CSS05,HK10,VK11}.

However, coarse-grained simulations show that the propensity to tilt is
also influenced by more generic factors. Venturoli \etal\ have reported that
cylindrical inclusions with larger radius tilt less than inclusions with small
radius \cite{VSS05}. Neder \etal\ have identified hydrophobicity as another
crucial factor determining tilt \cite{NNWS12}. In systematic studies of a variety
of simple inclusions with cylindrical shape and similar radii, embedded in a
model bilayer of the Lenz type, they found that the behavior of different
proteins mainly depended on their free energy of insertion, \ie, their binding
free energy. Weakly hydrophobic inclusions with negative binding free energies
(which stayed inside the membrane due to kinetic free energy barriers) react to
hydrophobic mismatch by tilting.  Strongly hydrophobic inclusions with binding
energies in excess of $100\,k_\romB T$ deform the membrane. For the probably most common
weakly bound inclusions with binding energies around $10\,k_\romB T$, the situation
is more complicated: upon increasing hydrophobic mismatch, inclusions first
distort the bilayer, and then switch to a tilted state once a critical mismatch
parameter is reached.  Tilting thus competes with the formation of
dynamic complexes consisting of proteins and a shell of surrounding, stretched
lipids, and the transition between these two states was found to be
discontinuous.

In the case where the membrane is deformed, the deformation profiles can be
compared to a variety of theories \cite{O78,O79,MB84,FB93, FBS95, JM04,BKM03}.
Both in coarse-grained \cite{VSS05,WBS09} and atomistic \cite{CP07}
simulations, it was reported that membrane thickness profiles as a function of
the distance to the protein are not strictly monotonic, but exhibit a weakly
oscillatory behavior. This feature is not compatible with membrane models that
predict an exponential decay \cite{O78,O79,JM04}, but it is nicely captured by
the coupled elastic monolayer models discussed earlier \cite{ABD96,BB06,WBS09}.
Coarse-grained simulations of the Lenz model showed that the coupled monolayer
models describe the profile data at a quantitative level, with almost no fit
parameters except the boundary conditions \cite{WBS09,NNWS12}. 

In membranes containing several inclusions, the membrane thickness deformations
induce effective interactions between inclusions. These have also been studied
within the Lenz model \cite{WBS09,NWN11} and other coarse-grained models
\cite{SGW08,MVS08}. The comparison with the elastic theory is less convincing,
due to the fact that many other factors such as local lipid packing contribute
to the effective potential of mean force, which cannot easily be separated from
the pure hydrophobic mismatch contribution \cite{WBS09}. Except for inclusions
with very large radii \cite{MVS08}, the hydrophobic mismatch contribution to
the effective interactions was generally found to be attractive.  

\subsection{Curvature mediated interactions between proteins}
\label{ssec:curvature}

\subsubsection{The mystery of the sign}

A very striking experimental demonstration of membrane curvature mediated
interactions was given by Koltover \etal\ in 1999
\cite{KoltoverRaedlerSafinya99}. These authors mixed micron-sized colloidal
particles with giant unilamellar vesicles to which they could adhere. While in
the absence of vesicles the colloidal particles showed no tendency to aggregate
in solution, they quickly did once they adsorbed onto the vesicles. Since it
was also evident from many micrographs that the colloids induced local bending
of the vesicle's membrane, the experiment strongly pointed towards membrane
curvature mediated attractions between the adhering colloids. This, however,
was very surprising: While interactions were indeed expected, the force should
have been \emph{repulsive}, as predicted six years earlier by Goulian \emph{et
al.}\  \cite{GouBruPinEPL}. Interestingly, the prefactor of this interactions
had to be corrected twice \cite{GouBruPinEPLe, FournierDommersnes97}, but this
did not change the outcome: the colloids should have repelled. It was soon
understood that objects that cause \emph{anisotropic} deformations could in
fact orient and then attract \cite{DommersnesFournier99, DommersnesFournier02,
Fournier-etal03}, but the colloids of Koltover \etal\ were isotropic (as far as
one could experimentally tell).

In what follows we will try to provide a glimpse into this mystery. A big part
of it has to do with too careless a use of the statement ``theory has
predicted''. Theory always deals with model systems and makes simplifying
assumptions, and this particular problem is fraught with seemingly
inconsequential details that could and sometimes do matter.

\subsubsection{The nonlinear ground-state--Take I}

The relevant field Hamiltonian pertaining to the curvature-mediated interaction problem is Eqn.~(\ref{eq:Helfrich-Hamiltonian})---minus several terms which will not matter. For a start, the last term involving the edge tension $\gamma$ does not arise in the absence of any membrane edge. The spontaneous bilayer curvature $K_0$ usually vanishes for symmetry reasons. If lipids can flip between the two leaflets, their chemical potential must be the same in both, and if no other symmetry-breaking field is present, this means that $K_0=0$. Unfortunately, membrane curvature \emph{itself} breaks the bilayer symmetry, and any existing lipid composition degree of freedom must couple to the geometry \cite{WuMcConnel75, Gebhardt-etal-77, Markin-81, Andelman-etal-92, Seifert-93, Cooke06}. So let us for now assume that this is not the case and take a note of this first nontrivial assumption. Moreover, in actual biomembranes none of this need be true since active and passive processes can maintain an asymmetric lipid composition across the two leaflets \cite{Bretscher-72, Daleke-07}. Finally, the term involving the Gaussian curvature can be dropped here, since we will neither encounter edges nor topology changes, and so the Gauss-Bonnet theorem will work in our favor.

What remains is the simpler Hamiltonian (\ref{eq:Monge-1}), but this looks quite formidable in Monge parametrization, as this very equation shows. To make any progress with something as forbidding as this appears quite unlikely. And yet, not all hope is lost. For a spherical particle attached to an asymptotically flat membrane the nonlinear shape equation has an exact solution, namely, a catenoid. This is an axisymmetric minimal surface with $K\equiv 0$ and hence obviously minimizes the left hand side of Eqn.~(\ref{eq:Monge-1}). If one adds additional lateral membrane tension, the exact shape of the membrane around a single adhering spherical particle can no longer be calculated analytically, but numerical solutions are relatively easy to come by using an angle-arclength parametrization \cite{Deserno-wrapping-04}. Unfortunately, we need to know the solution for \emph{two} particles, and in the absence of axisymmetry this is difficult---even numerically. It has been done \cite{ReynwarDeserno11}, but before we discuss this approach, let us first see what results we can analytically wrest from these equations.

Even for the full nonlinear problem the tight link between geometry and surface stress permits one to express mediated interactions as line integrals over the equilibrium membrane geometry. For instance, picture two spherical particles bound to a membrane, held at some mutual distance. If the particles are identical, then this will give rise to a mirror-symmetric membrane shape, and it may be shown that the force between these particles can be written as \cite{MDG-05a, MDG-05b}
\begin{equation}
F = \frac{1}{2}\kappa \int \romd s \; \left\{ K_{\perp}^2-K_{||}^2\right\} \ , \label{eq:curv-int-exact}
\end{equation}
where for simplicity we restricted to the tensionless case. The integral runs across the symmetry curve (the intersection of the membrane with the mirror plane), $K_{||}$ is the local curvature of that curve and $K_\perp$ the local curvature perpendicular to that curve. The sign convention is such that a negative sign implies attraction. To get an interaction strength out of Eqn.~(\ref{eq:curv-int-exact}) we need these curvatures, for which we need to solve the shape equations after all. Unfortunately, not even the sign of the interaction is evident form Eqn.~(\ref{eq:curv-int-exact}), since the \emph{difference} of two squares enters the integrals. Had we been curious instead about the interaction (per unit length) between two parallel rods on the membrane, we would have been in a better position: Now $K_{||}$ would be zero and the interaction would be clearly repulsive (even though we still don't quite know how strong it is). It seems that in order to make headway, we must solve the shape equation. The only hope to do this in reasonable generality using analytical tools is to linearize them.

\subsubsection{Linearization and superposition approximation}

Linearizing the nonlinear geometric functional means restricting to the first term in the integrand of Eqn.~(\ref{eq:Monge-2}). If we add a surface tension $\Gamma$, this means looking at the energy density $\frac{1}{2}\Gamma(\nabla h)^2+\frac{1}{2}\kappa(\Delta h)^2$, where $\nabla$ and $\Delta$ is the two-dimensional (flat!) surface gradient and Laplacian, respectively. A functional variation yields
\begin{equation}
\Big[\Gamma \Delta + \kappa \Delta\Delta \Big]\;h(\VECr) = 0 \ .
\end{equation}
This shape equation is of fourth order, but it is \emph{linear}. Unfortunately, in the present context we must solve it for a two-particle problem with finite-sized particles, and therein lies the rub: the operator in square brackets is not separable in any simple coordinate system, so we have to deal with the fact that this equation is indeed a \emph{partial} differential equation.

A popular trick to avoid this problem rests on the following reasoning: If the equation is linear, one might first want to look for a solution of the one-particle problem and then simply create the two-particle solution by superposition. We can then apply Eqn.~(\ref{eq:curv-int-exact}) to calculate the force, which in the present example would yield the interaction potential \cite{Deserno_BiomembraneFrontiers09}
\begin{equation}
U(r) = 2\pi\kappa\;\tilde{\alpha}^2\;{\rm K}_0(d/\lambda)
\qquad\text{with}\quad
\lambda = \sqrt{\frac{\kappa}{\Gamma}}
\;\;\;\;\text{and}\;\;
\tilde{\alpha}=\frac{\alpha}{{\rm K}_1(r_0/\lambda)} \ . \label{eq:superposition}
\end{equation}
Here, $r$ is the distance between the particles, $r_0$ is the radius of the circular contact line at which the membrane detaches from the colloid, $\alpha$ is the angle with respect to the horizontal at which it does so, and the ${\rm K}_\nu$ are a modified Bessel function of the second kind. This solution is analytical, simple, and wrong. Or more accurately, it only holds when $r\gg\lambda\gg r_0$, a restriction which excludes the interesting tensionless limit in which $\lambda\rightarrow\infty$. The \emph{mathematical} reason is that superposition in the way celebrated here is not allowed: yes, superpositions of solutions to linear equations are still solutions, but superpositions of solutions, each of which only satisfies some part of all pertinent boundary conditions, generally do not satisfy \emph{any} boundary condition and are thus not the solutions we are looking for. The \emph{physical} reason why the superposition ansatz in this case fails is because the presence of one colloid on the membrane which creates a local dimple will abet a nearby colloid to \emph{tilt}, thereby changing the way in which that second colloid interacts with the membrane and, in turn, the first one.

\subsubsection{Linearization and a full two-center solution}
 
One way to circumvent the superposition approximation is to solve the full two-center problem. This is of course much more tedious, and in fact can only be handled as a series expansion (in which one satisfies the boundary conditions at both particles up to some order in the multipoles and an expansion in the smallness parameter $r_0/r$. This calculation has been done by Weikl \emph{et al.}\ 
\cite{WeiklKozlovHelfrich98}, leading to
\begin{equation}
U(r) = 2\pi\kappa\left(\frac{\alpha r_0}{\lambda}\right)^2\left\{{\rm K}_0(r/\lambda)+\left(\frac{r_0}{\lambda}\right)^2{\rm K}_2^2(r/\lambda) + \cdots\right\} \ .\label{eq:WKH}
\end{equation}
Notice that in the case $r\gg\lambda\gg r_0$ this indeed reduces to Eqn.~(\ref{eq:superposition}), while in the more interesting limit in which the tension vanishes it reduces to
\begin{equation}
U(r) = 8\pi\kappa\,\alpha^2\,\left(\frac{r_0}{r}\right)^4 \ , \label{eq:GBP}
\end{equation}
which is indeed the solution of Goulian \emph{et al.} \cite{GouBruPinEPL}, amended by the prefactor corrections \cite{GouBruPinEPLe, FournierDommersnes97}. In fact, these authors have actually written down the solution for the case of two non-identical particles 1 and 2 with detachment angles $\alpha_1$ and $\alpha_2$. If we also make their radii $r_i$ different, we find \cite{YolDes-membrane}
\begin{equation}
U(r) = 4\pi\kappa (\alpha_1^2+\alpha_2^2)\frac{r_1^2r_2^2}{r^4} \ . \label{eq:GBP12}
\end{equation}
Notice that unlike what one might have guessed from Eqn.~(\ref{eq:GBP}) the potential (and thus the force) is \emph{not} proportional to the product of the two detachment angles. The actual form of the prefactor, $\alpha_1^2+\alpha_2^2$, is highly suggestive of an entirely different underlying physics, as we will now see.

\subsubsection{Linearization using effective field theory}

Eqns.~(\ref{eq:WKH}), (\ref{eq:GBP}) and (\ref{eq:GBP12}) are expansions of the exact solution for large distance. Working out higher order terms appears reasonably forbidding, given that one has to push a difficult multi-center problem to high order. However, there is a way to disentangle the multi-center problem from the interaction problem.

We have seen that the \emph{physical} reason why the superposition approximation fails is the induced tilting of neighboring colloids. More generally, any finite particle in contact with the membrane will induce extra membrane deformations if the membrane in its vicinity is perturbed. This is simply a polarization effect: Any ``incoming'' field interacts with the boundary conditions imposed by the particle and these then create new ``outgoing'' fields. Superposition of fields would work for point particles, but these don't capture these polarization effects, \emph{unless} we equip them with the requisite \emph{polarizabilities}. But this of course we can do. We can write a new Hamiltonian of interacting point particles, where each of them has the same polarizabilities as the actual finite size particles of the situation we actually wish to describe. This works by adding terms to the Hamiltonian which are localized at the position of the particle and which couple to the field in the same way that a local polarizability would. For instance, if a particle at the position $r_\alpha$ has a dipole polarizability $C_\alpha^{(1)}$, we must add the term $\frac{1}{2}C_\alpha^{(1)}[h_i(\VECr_\alpha)]^2$
to the Hamiltonian, where the index $i$ is again a derivative. The energy increases quadratically with the gradient of the local field---exactly as for a dipole polarizability. The only remaining question is: where do we get the polarizabilities from? Answer: just like in classical electrostatics, by calculating the response of \emph{one particle} in a suitably chosen external field and comparing the full theory with the effective point particle theory.

This idea is an example of what is referred to as \emph{effective field theory} \cite{TASI}, and it has been used for a host of vastly diverse problems, ranging from black holes in general relativity \cite{GoldbergerRothstein06, Porto-etal-11} to finite-size radiation corrections in electrodynamics \cite{Galley-etal-10}. The first application in the context of fluid soft surfaces was given by Yolcu \etal\ \cite{YolRotDesEPL, YolRotDes-longfilm}. For two axisymmetric particles on a membrane Yolcu and Deserno showed that Eqn.~(\ref{eq:GBP12}) extends as follows \cite{YolDes-membrane}:
\begin{equation}
U(r) = 4\pi\kappa (\alpha_1^2+\alpha_2^2)\frac{r_1^2r_2^2}{r^4} + 8\pi\kappa\left(\frac{\alpha_1}{r_1}-\frac{\alpha_2}{r_2}\right)^2\frac{r_1^4r_2^4}{r^6} + \cdots
\end{equation}
Notice that the next order correction is also repulsive and in fact vanishes for identical particles (in contrast to some earlier calculations \cite{DomFouEPL} which missed terms that contribute at the same order).

\subsubsection{Fluctuation mediated interactions}

It has long been known that even two \emph{flat} circular particles on a membrane feel an interaction, since their boundaries affect the fluctuation spectrum of the membrane and thus its free energy. These forces are proportional to the thermal energy $k_\romB T$ and not to the surface rigidity $\kappa$ and are examples of \emph{Casimir interactions} in soft matter systems \cite{GolestanianKardar99}. For circular discs on a tensionless membrane they are attractive and, to lowest order, decay like the $4^{\rm th}$ power of distance \cite{GouBruPinEPL, ParLub96, DomFouEPL, HelfrichWeikl01}.

The true beauty of the effective field theory approach described in the previous section is that it also greatly simplifies force calculations on \emph{thermally fluctuating} surfaces \cite{YolRotDesEPL, YolRotDes-longfilm, YolDes-membrane}. For two flat rigid particles of radii $r_1$ and $r_2$ Yolcu and Deserno find \cite{YolDes-membrane}
\begin{equation}
-\frac{U(r)}{k_\romB T} = 6\,\frac{r_1^2r_2^2}{r^4} +10\,\frac{r_1^2r_2^4+r_1^4r_2^2}{r^6} +3\,\frac{r_1^2r_2^2(5r_1^4+18r_1^2r_2^2+5r_2^4)}{r^8} + \cdots \ . \label{eq:mem-Casimir}
\end{equation}
The leading order is well known,\footnote{Unfortunately, in the first paper which discusses this force, Goulian \etal\ \cite{GouBruPinEPL} claim that the prefactor is 12, a mistake that is not fixed during the prefactor-fixing in \cite{GouBruPinEPLe}.} all higher orders are new. In fact, if one restricts to \emph{identical} particles, many more orders can be readily written down:
\begin{equation}
-\frac{U(r)}{k_\romB T} = \frac{6}{x^4} + \frac{20}{x^6}+\frac{84}{x^8}+\frac{344}{x^{10}}+\frac{1388}{x^{12}} + \frac{5472}{x^{14}} +\frac{21370}{x^{16}} + \frac{249968}{x^{18}} \cdots \ ,
\end{equation}
where $x=r/r_0$.

So here we have the first example of an attraction. Could these forces explain the aggregation observed by Koltover \etal\ \cite{KoltoverRaedlerSafinya99}? This is difficult to say. First, in the case of almost flat membranes, which all these calculations implicitly assume by using linearized Monge gauge, the ground state repulsion (\ref{eq:GBP12}) overwhelms the fluctuation contribution (\ref{eq:mem-Casimir}) once $\alpha>\alpha_{\rm c}=\sqrt{3k_\romB T/4\pi\kappa}$, and for a typical choice of $\kappa=20\,k_\romB T$ this gives the rather small angle $\alpha_{\rm c}\approx 6^\circ$. Most likely the colloids in the experiments by Koltver \etal\ imposed much bigger deformations, but it is hard to say what happens to both forces at larger angles. In the next section we discuss the numerical solution of the ground state problem, but at present no calculations exist which push the Casimir force beyond the linear regime, except in the case of two parallel \emph{cylinders}, for which Gosselin et al. find, rather remarkably, that the Casimir force is \emph{repulsive} \cite{Gosselin2012}.

\subsubsection{The nonlinear ground-state--Take II}

The various linear calculations show that two axisymmetric colloids on a membrane should repel. But as the detachment angles $\alpha_i$ increase, it becomes harder to justify the linearization. The expansion in Eqn.~(\ref{eq:Monge-2}) ultimately rests on the smallness of $|\nabla h|$, an expression that should be compared to $\tan\alpha_i$. But once higher order terms matter, Monge parametrization not only becomes technically impenetrable; it is even incapable of dealing with membrane shapes that display \emph{overhangs}. It is hence preferable to discard it in favor of a more general numerical surface triangulation.

Reynwar and Deserno \cite{ReynwarDeserno11} have studied the interaction problem for identical axisymmetric colloids with large angles $\alpha_i$, using the package ``Surface Evolver'' by Brakke \cite{Brakke92}. For small angles $\alpha_i$ the large distance predictions coincide well with Eqn.~(\ref{eq:GBP}), but they break down rather abruptly as soon as $r<2r_0$, which is when the particles would touch \emph{unless} they could also tilt out of each other's way. For large $\alpha_i$ the linear predictions substantially overestimate the repulsion. Interestingly, for the special case $\alpha=\pi/2$ the repulsive force goes through a maximum (around $r/r_0\approx 1.8$), and it \emph{decreases} upon moving the particles even close together until it vanishes at $r/r_0\approx 1$. At even closer distances the particles \emph{attract}. Attractive forces must exist also for detachment angles smaller than $\pi/2$, but Ref.~\cite{ReynwarDeserno11} does not attempt to find the minimal angle at which this happens. They certainly also exist for angles bigger than $\pi/2$, even though it might be that there is also a largest angle for which they exist. In any case, only for $\alpha=\pi/2$ does the attraction persist all the way to $r=0$.

A simple close distance approximation can be devised to understand the necessity of a sign-flip. At sufficiently close distances the two particles tilt so much that they almost face each other, and the membrane between them assumes a shape similar to a cylinder, which 
is capable of transmitting tensile forces as we have seen in Sec.~\ref{ssec:material-parameters}. For angles close to $\pi/2$ this theory suggests \cite{ReynwarDeserno11}
\begin{equation}
\frac{Fr_0}{\pi\kappa} = \frac{1}{x^2}+\frac{1-\sin\alpha}{x} - 1 + \mathcal{O}(x)
\qquad\text{with}\quad
x = \frac{r}{2r_0\,\cos\alpha} \ .
\end{equation}
Observe that the first two terms vanish for $\alpha=\pi/2$, which leaves the (attractive) force $F=\pi\kappa/r_0$, which is \emph{half} the value transmitted through a cylindrical membrane tube---see Eqn.~(\ref{eq:tether-force}). The missing factor of $2$ derives from the fact that this calculation is not done at constant area but at constant (in fact: zero) tension. The numerical calculations suggest that indeed $F(r)$ approaches a constant as $r\rightarrow 0$, even though it seems slightly off from the expected value $-\pi\kappa/r_0$.

\subsubsection{Curvature mediated interactions in simulations}

The experiments by Koltover \etal\ claim that isotropic colloids on membranes experience a surface (presumably: curvature-) mediated attraction. All theories we have discussed so far claim the force is repulsive, unless one goes to pretty large detachment angles. Can simulations shed more light onto the problem? If so, it will not be necessary to represent the bilayer in any greater detail, since only fluid curvature elasticity needs to be captured.

Reynwar \etal\ have investigated this problem using the Cooke model, amended by simple generic particles with some given isotropic curvature \cite{Reynwar07}. They showed that indeed strongly mem\-bra\-ne-de\-for\-ming colloids experience attractive pair interactions. Subsequent more detailed studies revealed that these are compatible with the numerical results discussed in the previous section \cite{ReynwarDeserno11}. However, they also showed that a large number of \emph{weakly} membrane deforming colloids still aggregate---in fact, that they can drive vesiculation of the membrane \cite{Reynwar07}. This is surprising, since these particles exhibited detachment angles at which the ground state theory clearly insists on a repulsive pair potential.

However, just because the pair potentials are repulsive does not yet prove that aggregation cannot happen, since curvature mediated interactions are not pairwise additive, as first pointed out by Kim \etal\ \cite{Kim98, KimNeuOster99}. These authors provide a general formula for an $N$-body interaction, and even though it is really only accurate up to the triplet level \cite{YolDes-membrane}, it does show that the contributions beyond pairs can lower the overall repulsive energy; for instance, they show that certain multi-particle configurations are indeed marginally stable instead of being driven apart. In a later publication Kim \etal\ \cite{KimChouRudnick08} show that an infinite number of periodic lattices exists for which summing the non-pairwise interactions preserve zero membrane bending energy. Again, since their non-pairwise form is only accurate up to triplet order, it is not clear whether this result remains true if \emph{all} orders are considered. M\"uller and Deserno have alternatively treated this problem using a cell model \cite{MullerDeserno10}, in which a regular lattice of particles is replaced by a single particle within a cell, plus boundary conditions that mimic the presence of other surrounding particles. They prove that within this approximation the lateral pressure between colloids is \emph{always} repulsive, even in the nonlinear regime;\footnote{They used the same techniques that also led to the exact Eqn.~(\ref{eq:curv-int-exact}), only that in the cell model case the sign is evident from the expression.} how well the cell model actually captures a multi-particle assembly is difficult to say, though. Auth and Gompper have also used a cell model approach \cite{AuthGompper09}, but they specifically apply it to a \emph{curved} background membrane. They argue that even if the forces are repulsive, they might be \emph{less} repulsive---and thus the free energy per colloid smaller---if the background membrane is curved, since this background curvature screens the repulsion between the colloids. This could provide a driving force for creating curved vesicle buds from flat membranes studded with isotropic membrane curving colloids, provided the average area density of colloids remains fixed. The latter is usually the case in simulations, and Auth and Gompper show that the sizes of the vesicle which detach from the parent membrane for differently curved colloids is compatible with what Reynwar \etal\ \cite{Reynwar07} observe. What would fix this density in real systems is less clear, but it is conceivable that this is yet another situations where rafts come into play: If the membrane-curving particles have to stay within a finite raft, their mutual repulsion can, by virtue of the mechanism discussed by Auth and Gompper, lead to a budding of that raft domain.

\vspace*{0.3em}

In conclusion we see that the situation is substantially more tricky than the seemingly simple questions ``do membrane curving particles attract or repel?'' leads one to expect. Nonlinearities, multibody interactions, fluctuations, background curvature, boundary conditions, anisotropies, are only some of the ``details'' which affect the answer to this question. At the moment the situation remains not completely solved, but the results outlined in this section should provide a reliable guide for future work.

\section{Multiscale modeling of lipid and membrane protein systems}

\subsection{Multiscale modeling: approaches and challenges}

As we have seen in the previous sections, coarse grained lipid models have been enormously successful at investigating phenomena in lipid bilayers and lipid bilayer/protein systems. In particular,
rather coarse, generic models that reduce the lipids to their most essential features and shed almost all chemical specificity have enormously contributed to our understanding of effective interactions, generalized processes, and their driving forces.
A different branch of coarse grained models, the already mentioned bottom-up models, has progressed quite dramatically in the past decade as well. These models are not developed as stand-alone models with parameters derived to reproduce some desired experimentally known feature of the system. They are developed in a bottom-up way with the help of an underlying higher-resolution (atomistic) model. Therefore, frequently the terms ``multiscale modeling'' or ``systematic coarse graining'' are used. 
These models allow to stay closer to an atomistic system and to retain more chemical specificity, and due to their bottom-up construction they offer the opportunity to go back and forth between a coarse grained and an atomistic level of resolution using so-called backmapping techniques. 

It should be noted though, that this closeness between levels of resolution does come at a cost: upon reducing the number of degrees of freedom the models become strongly state point dependent and it necessarily becomes impossible to accurately represent {\em all} properties of the underlying atomistic system with the coarse grained model. In particular the representation of thermodynamic as well as structural properties is a severe challenge that has been subject of a multitude of studies over the last years \cite{John07.1}. The question of representability and the unavoidable choice of parametrization target properties that has to be made has led to a number of different systematic coarse graining approaches which are often divided into two general categories: {\em (i)} methods where the CG parameters are refined so that the system displays a certain thermodynamic behavior (typically termed {\em thermodynamics-based}) 
\cite{NIEL03.1,Marrink04,marrink_martini_2007,MOGN08.1,DEVA09.1}
or  {\em (ii)} methods where the CG system aims at reproducing the configurational phase space sampled by an atomistic reference system (often misleadingly termed {\em structure-based}) \cite{TSCH98.1,LYUB95.1,FMP-review02,REIT03.1,PETE08.1,MURT09.1,LYUB10.1,SAVE09.1,MEGA11.1,MUKH12.1,
 IzvekovVoth05,MSC-I-08, SHEL08.1}. 
Representability limitations lead to the observation that a structure-based approach does not necessarily yield correct thermodynamic properties such as solvation free energies or partitioning data while  thermodynamics-based potentials may not reproduce microscopic structural data such as the local packing or the structure of solvation shells. Closely related  are also the inevitable transferability problems of CG models: all CG models (in fact also all classical
atomistic force fields) are state-point dependent and cannot
necessarily be -- without reparametrization -- transferred to
different thermodynamic conditions (temperature, density, concentration, system composition, phase, etc.) or a different chemical or molecular environment (e.g. a certain chemical unit being part of different macromolecular chains).
Structural and thermodynamic representability and
state-point transferability questions are often intimately linked,
since the response to a change in state point corresponds to
representing certain thermodynamic properties.
Intensive research
is currently devoted to this
problem \cite{SILB06.1,ALLE09.1,MULL09.1,VILL10.1,IZVE10.1,SHEN11.1,BRIN11.1,MEGA11.1,MUKH12.1}, since the understanding of potential and limitation of coarse grained models is a necessary prerequisite to applying them to complex biomolecular problems and systems such as multi-protein complexes in biomembranes for the following reason:
CG models are usually developed based on smaller less complex reference systems -- a reference simulation of the actual target system is by construction prohibitive, otherwise the whole coarse graining effort would not be necessary in the first place.
Consequently, it is essential to understand transferability among different concentrations, compositions and environments to be able to put these subsystem-based models together and obtain a reliable model for the actual -- more complex -- target system. 
In the following we will show for one example -- the  light harvesting complex of green plants (LHCII) -- 
some aspects  of multiscale modeling of  membrane protein systems and some of the  problems that  need to be addressed 
if one wants to go beyond generic coarse grained models and retain a certain level of chemical specificity.

\subsection{The light harvesting complex}

The major light-harvesting complex (LHCII) of the photosynthetic apparatus in green plants binds more than half of the plant's chlorophyll (Chl) and is presumably the most abundant membrane protein on Earth. It has become an intensely studied model membrane protein for several reasons. Its structure is known in near-atomic detail \cite{Liu.2004, Standfuss.2005}, and much of its biochemistry has been elaborated in the past decades \cite{Schmid.2008}. Moreover, LHCII spontaneously self-organizes from its protein and pigment components in vitro; therefore, recombinant versions of it can easily be produced and modified almost at will \cite{Yang.2003}. The asssembly of LHCII and the concomitant folding of its apoprotein has been studied in some detail \cite{Horn.2007, Dockter.2009}. Both processes occur spontaneously upon combining the unfolded apoprotein and pigments in detergent solution. In vivo, the assembly of LHCII takes place in the lipid environment of the thylakoid membrane and, most likely, is influenced by the lipid and protein components of this membrane. This is difficult to analyse experimentally since, so far, the self-organisation of LHCII cannot be achieved yet in a lipid membrane environment. Recently, also the disassembly of LHCII and the role of the bound/dissociating pigments in the falling apart of LHCII trimers has become subject of increased interest. These pigments constitute about 1/3 of the total mass of LHCII and, according to the structure, significantly contribute to the stability of the pigment-protein complex. The structural behavior of LHCII has been analyzed by circular dichroism (CD), fluorescence, and electron paramagnetic resonance (EPR) \cite{Yang.2003, Horn.2004, Dockter.2009, Dockter.2012}.

One important aspect of LHCII that specifically relates to other aspects discussed in the present review is the question of how the membrane environment (lipid composition, membrane curvature, etc.) affects the association of LHCII monomers to form trimers and the assembly of these trimers into the antenna complex around the photosynthetic reaction centers. The non-bilayer forming lipid MGDG constitutes half of the thylakoid membrane. This membrane maintains its lamellar structure only with proteins inserted, predominantly LHCII which, due to its concave shape, eases the curvature pressure exerted by MGDG. It has been suggested that this curvature pressure is a driving force for protein interaction in the membrane \cite{Garab.2000}, however, since it is not known whether, e.g., the formation of supercomplexes of LHCII trimers eases or increases curvature pressure, it is unclear whether MGDG (or other curvature pressure-increasing lipid components) promote or inhibit the formation of such supercomplexes. Likewise, the composition of the lipid membrane and the membrane properties such as its curvature pressure most likely influence the folding of the LHCII apoprotein and its assembly with pigments.

LHCII commends itself as a useful model to study the the influence of the lipid membrane on the assembly and structural behavior of membrane proteins in general because of its known structure, its availability in a recombinant form, and its self-organisation, at least in detergent micelles. Moreover, the Chl molecules bound serve as built-in fluorescence markers for monitoring the structural behavior of the pigment-protein complex. To be able to correlate experimental observations of aggregate formation with predictions from theory, recombinant LHCII has been inserted in liposomes and assayed for complex-complex distances by inter-complex FRET measurements and for aggregate formation by quantitating aggregate-induced fluorescence quenching (data to be published). Moreover, to test the simulation of pigment-protein assembly in the membrane environment, procedures are being established to dissociate and re-associate recombinant LHCII in liposomes and to use an in-vitro expression system to insert the protein into liposome membranes \cite{Yildiz.2012}.

A multiscale simulation model to study the LHCII complex requires as a first step model parameters for all components involved. As already mentioned  above, it will be neither possible nor useful to parameterize a CG model based on the actual multicomponent (lipid bilayer/protein/pigments) system but one would rather develop models for sensibly chosen subsystems. While typically parameters for the protein and the lipid bilayer can be found in many standard forcefields, a challenging first task is to obtain a reliable model for the pigments -- irrespective of the level of resolution. 
For many biological applications the MARTINI CG forcefield -- that has already been described above -- has become very popular and successful, in particular for lipid bilayer and protein systems. To employ the MARTINI forcefield for simulations of the pigmented LHCII, a CG description and model parameters for the pigment molecules needs to be added. We have developed a coarse-grained model of the chlorophyll pigments (Chl {\it b} and Chl {\it a}) which can be embedded into the existing MARTINI force field to study the pigmented LHCII trimer in the future. To do this, Chl {\it b} and Chl {\it a} were parameterized in the presence of the lipid bilayer. This reference system for parametrization was chosen for two reasons:  most importantly, the Chl-lipid interactions are highly relevant for the formation and behavior of the LHCII protein-pigment complex in the lipid bilayer. 50\% of the pigment molecules in the plant are bound to the light harvesting complex, with 42 Chl molecules per LHCII trimer. In vitro studies have shown that the folding of the LHCII apoprotein and the pigment binding to the protein are tightly coupled processes. In the LHCII monomer, many Chl pigments are situated in the outer region of the protein, effectively forming an interface between protein and lipids. Consequently the Chl-lipid interactions are most likely important for the assembly and stability of the trimer. A second reason for choosing the Chl-lipid  system as reference  for which the interactions between the MARTINI standard forcefield and Chl can be tuned  is that it is more tractable  compared to the fully pigmented LHCII membrane protein complex.
The CG model for Chl {\it b} and {\it a} in the DPPC bilayer was derived based on a combination of a structure-based approach for bonded and a mixed structure-based and partitioning-based approach for non-bonded interaction potentials to fit the thermodynamics-based MARTINI force field. The CG model for Chl molecules  follows the degree of coarse graining of the MARTINI forcefield. Somewhat in line with the general MARTINI parameterization philosophy, which focuses on partitioning properties, the non-bonded parameters were chosen such that the distribution of the CG Chl beads between hydrophilic and hydrophobic regions in the bilayer is correctly represented -- compared to the atomistic
reference simulation. 
Here, particular attention was paid to the interactions of the polar center of the porphyrin ring with the lipid beads and to the polarity of the aromatic ring which needs to be carefully tuned to obtain the correct distribution between the polar headgroup  and the hydrophobic tail regions of the lipid bilayer.
The bonded interactions in the CG pigments were derived such that the
coarse grained model reproduces the shape and the conformational behaviour of the atomistic Chl molecules -- the
overall shape of the porphyrin ring and the different conformations of the phytol 
tail are well represented in this CG model. As a last aspect of validation of the CG model we have analyzed the propensity of the Chl pigments to aggregate in the lipid bilayer: It was found that Chl molecules do aggregate, with clusters that from and break multiple times in the course of the simulation, i.e. the aggregation is not overly strong. Qualitatively, these data are corroborated by fluorescence quenching experiments which show that chlorophylls in lipid bilayers have a tendency to aggregate at low lipid:Chl ratios of less than 1250 lipids/chlorophyll.
Summarizing, the structural behavior, the distribution of the pigments in the bilayer (which are indicative of a correct balance of hydrophobicity/hydrophilicity) and the pigment association is very well represented in the CG model compared to atomistic simulations and experimental data. \cite{Debnath.submitted}.

\begin{figure}
\centering
\includegraphics[width = 1\textwidth]{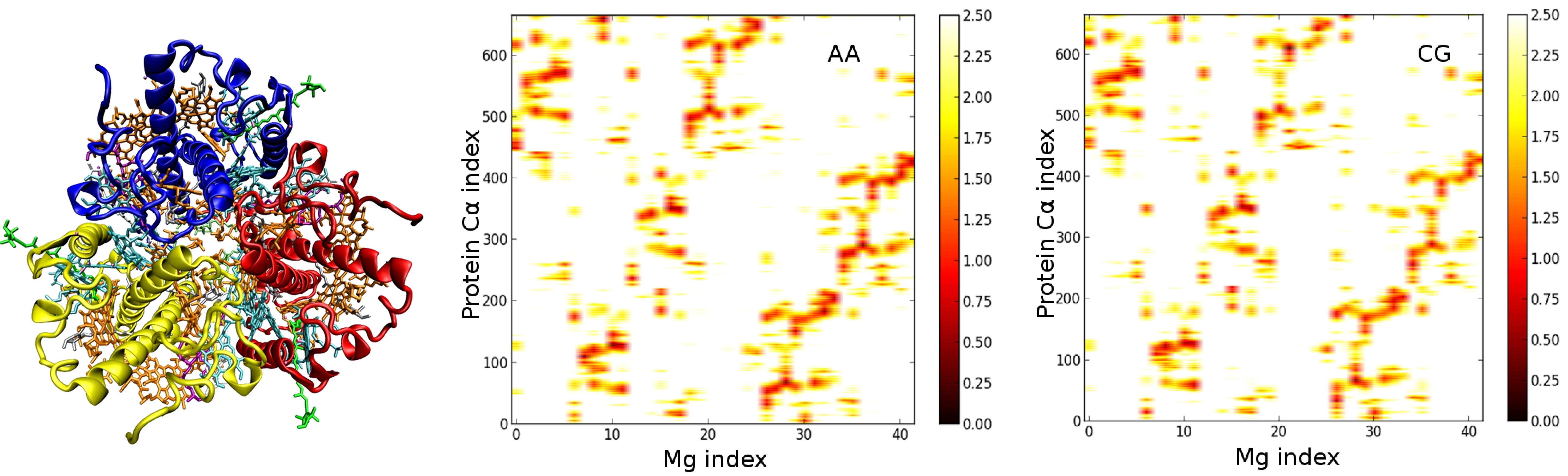} 
\caption{Left panel: Top view of an LHCII trimer (colors according to chain or molecule type: blue - chain A, red - chain B, green - chain C,
cyan - Chl {\it b}, pink - Chl {\it a}). Middle and right panels: Contact maps
between Chl pigments and protein residues of LHCII trimer Ð drawn as distance
maps between the C$\alpha$ atoms of the proteins ($y$-axis) and the Mg atoms of all Chl
pigments ($x$-axis) within a 2.5 nm cut-off for 70 ns long atomistic (middle panel) and 100
ns long CG (right panel) simulations. The maps show that the pigments are stably located in their binding sites for both levels of resolution. }
\label{cg-trimer}
\end{figure}

After driving the CG model parameters for the Chl-lipid system, this new model was now combined with the MARTINI model for proteins to perform some first simulations of the pigmented LHCII complex (in trimeric as well as in monomeric form).
In addition classical atomistic (explicit solvent) simulations of trimeric and monomeric LHCII in a model membrane have been performed to provide a reference for validation of the CG simulations.
The first CG simulations of the LHCII complex have proven to be very promising. Unlike our initial attempts without the careful parameterization of the pigments, the trimeric protein-pigment complex has been structurally stable, most notably without the presence of any artificial elastic network between the protein core and the pigments (see Figure~\ref{cg-trimer}). The properties of the complex from the CG model are in excellent agreement with the atomistic ones. 
In the future, this CG model  will be used to study various aspects of LHCII protein/protein interactions in the lipid bilayer that  on the one hand go beyond the  time- and length scales accessible to atomistic simulations alone and on the other hand 
require a more chemically realistic description of the protein/pigment/lipid system than in typical generic CG models.

\clearpage

\section{Conclusions}

In the present chapter we have presented an overview of different approaches to study lipid membranes and membrane protein systems. We have reviewed theoretical and simulation approaches, and shown how generic lipid simulation models can be used to understand the principles that determine properties of lipid bilayers such as bending and Gaussian curvature modulus,  membrane tension, or fundamental phenomena such as the formation of  lipid rafts, or the curvature mediated interactions between proteins. 
In the concluding section it was  outlined how multiscale modeling can in principle go a step further by ensuring a certain chemical specificity while still benefiting from the time- and length-scale advantages of coarse grained simulations -- noting though that there are still a number of challenges in the area of systematic coarse graining that need to be addressed to be able to study complex multicomponent systems such as the the light harvesting complex of green plants. For this system  we have shown first steps toward a multiscale simulation model that allows to go back and forth between a coarse grained and an atomistic level of resolution and therefore permits immediate comparison to atomic level experimental data.

\begin{acknowledgement}
We would like to thank the many coworkers and colleagues who have contributed to theÊresearch about which we have reported here, in particular Ira Cooke, Jemal Guven, Vagelis Harmandaris, Gregoria Illya, Martin M\"uller, Benedict Reynwar, Ira Rothstein, Cem Yolcu, Frank Brown, ÊOlaf Lenz, Sebastian Meinhardt, Peter Nielaba, Beate West, Ananya Debnath, Christoph Globisch, Christoph Junghans, Shahoua Ding, Sabine Wiegand, Sandra Ritz, and Eva Sinner.

\end{acknowledgement}
%


\begin{thebibliography}{100}
\providecommand{\url}[1]{{#1}}
\providecommand{\urlprefix}{URL }
\expandafter\ifx\csname urlstyle\endcsname\relax
  \providecommand{\doi}[1]{DOI~\discretionary{}{}{}#1}\else
  \providecommand{\doi}{DOI~\discretionary{}{}{}\begingroup
  \urlstyle{rm}\Url}\fi

\bibitem{Ahmed-etal-97}
Ahmed, S.N., Brown, D.A., London, E.: {On the origin of
  sphingolipid/cholesterol-rich detergent-insoluble cell membranes:
  Physiological concentrations of cholesterol and sphingolipid induce formation
  of a detergent-insoluble, liquid-ordered lipid phase in model membranes}.
\newblock {Biochem.} \textbf{{36}}({36}), {10,944--10,953} ({1997})

\bibitem{ALLE09.1}
Allen, E.C., Rutledge, G.C.: {Evaluating the transferability of coarse-grained,
  density-dependent implicit solvent models to mixtures and chains}.
\newblock J. Chem. Phys. \textbf{130}, 034,904 (2009)

\bibitem{Andelman-etal-92}
Andelman, D., Kawakatsu, T., Kawasaki, K.: Equilibrium shape of two-component
  unilamellar membranes and vesicles.
\newblock Europhys. Lett. \textbf{19}(1), 57--62 (1992)

\bibitem{Antonny06}
Antonny, B.: Membrane deformation by protein coats.
\newblock Curr. Opin. Cell. Biol. \textbf{18}(4), 386--394 (2006)

\bibitem{ABD96}
Aranda-Espinoza, H., Berman, A., Dan, N., Pincus, P., Safran, S.: Interaction
  between inclusions embedded in membranes.
\newblock Biophys. J. \textbf{71}, 648 (1996)

\bibitem{Arkhipov08}
Arkhipov, A., Yin, Y., Schulten, K.: Four-scale description of membrane
  sculpting by bar domains.
\newblock {B}iophys. {J}. \textbf{95}(6), 2806--2821 (2008)

\bibitem{AtzbergerKramerPeskin07}
Atzberger, P.J., Kramer, P.R., Peskin, C.S.: {A stochastic immersed boundary
  method for fluid-structure dynamics at microscopic length scales}.
\newblock {J. Comput. Phys.} \textbf{{224}}({2}), {1255--1292} ({2007})

\bibitem{AuthGompper09}
Auth, T., Gompper, G.: Budding and vesiculation induced by conical membrane
  inclusions.
\newblock Phys. Rev. E \textbf{80}, 031,901 (2009).
\newblock \doi{10.1103/PhysRevE.80.031901}

\bibitem{Ayton02}
Ayton, G., Voth, G.A.: Bridging microscopic and mesoscopic simulations of lipid
  bilayers.
\newblock {B}iophys. {J}. \textbf{83}(6), 3357--3370 (2002)

\bibitem{Baumgart05}
Baumgart, T., Das, S., Webb, W.W., Jenkins, J.T.: Membrane elasticity in giant
  vesicles with fluid phase coexistence.
\newblock {B}iophys. {J}. \textbf{89}(2), 1067--1080 (2005)

\bibitem{BS12}
Benjamini, A., Smit, B.: Robust driving forces for transmembrane helix packing.
\newblock Biophys. J. \textbf{103}, 1227--1235 (2012)

\bibitem{BennunHoopesXingFaller09}
Bennun, S.V., Hoopes, M.I., Xing, C., Faller, R.: {Coarse-grained modeling of
  lipids}.
\newblock {Chem. Phys. Lipids} \textbf{{159}}({2}), {59--66} ({2009})

\bibitem{Berkowitz09}
Berkowitz, M.L.: {Detailed molecular dynamics simulations of model biological
  membranes containing cholesterol}.
\newblock {Biochim. Biophys. Acta -- Biomembranes} \textbf{{1788}}({1}),
  {86--96} ({2009})

\bibitem{BEM91}
Bloom, M., Evans, E., Mouritsen, O.G.: Physical properties of the fluid
  lipid-bilayer component of cell membranes: a perspective.
\newblock Quart. Rev. Biophys. \textbf{24}(3), 293--297 (1991)

\bibitem{Bo89}
Bo, L., Waugh, R.E.: Determination of bilayer membrane bending stiffness by
  tether formation from giant, thin-walled vesicles.
\newblock {B}iophys. {J}. \textbf{55}(3), 509--517 (1989)

\bibitem{BKM03}
Bohinc, K., Kralj-Iglic, V., May, S.: Interaction between two cylindrical
  inclusions in a symmetric lipid bilayer.
\newblock J. Chem. Phys. \textbf{119}, 7435--7444 (2003)

\bibitem{BHS06}
Botelho, A.V., Huber, T., Sakmar, T.P., Brown, M.F.: Curvature and hydrophobic
  forces drive oligomerization and modulate activity of rhodopsin in membranes.
\newblock Biophys. J. \textbf{91}, 4464--4477 (2006)

\bibitem{Brakke92}
Brakke, K.A.: The surface evolver.
\newblock Exp. Math. \textbf{1}, 141--165 (1992)

\bibitem{BBS11}
Brandt, E.G., Braun, A.R., Sachs, J.N., Nagle, J.F., Edholm, O.: Interpretation
  of fluctuation spectra in lipid bilayer simulations.
\newblock Biophys. J. \textbf{100}, 2104--2111 (2011)

\bibitem{BE10}
Brandt, E.G., Edholm, O.: Stretched exponential dynamics in lipid bilayer
  simulations.
\newblock J. Chem. Phys. \textbf{133}, 115,101 (2010)

\bibitem{BB06}
Brannigan, G., Brown, F.: A consistent model for thermal fluctuations and
  protein-induced deformations in lipid bilayer.
\newblock Biophys. J. \textbf{90}, 1501 (2006)

\bibitem{BB07}
Brannigan, G., Brown, F.L.H.: Contributions of gaussian curvature and
  nonconstant lipid volume to protein deformation of lipid bilayers.
\newblock Biophys. J. \textbf{92}, 864--876 (2007)

\bibitem{Brannigan-review06}
Brannigan, G., Lin, L., Brown, F.: {Implicit solvent simulation models for
  biomembranes}.
\newblock {Eur. Biophys. J.} \textbf{{35}}({2}), {104--124} ({2006})

\bibitem{Brazovskii75}
Brazovskii, S.A.: Phase transitions of an isotropic system to a nonuniform
  state.
\newblock Soviet Physics JETP \textbf{41}, 85--89 (1975)

\bibitem{Bretscher-72}
Bretscher, M.S.: {Asymmetrical Lipid Bilayer Structure for Biological
  Membranes}.
\newblock {Nature} \textbf{{236}}({61}), {11--12} ({1972})

\bibitem{BPS09}
Brewster, R., Pincus P. A.~Safran, S.A.: Hybrid lipids as a biological
  surface-active component.
\newblock Biophys. J. \textbf{97}, 1087--1094 (2009)

\bibitem{BRIN11.1}
Brini, E., Marcon, V., van~der Vegt, N.F.A.: {Conditional reversible work
  method for molecular coarse graining applications}.
\newblock Phys. Chem. Chem. Phys. \textbf{13}, 10,468--10,474 (2011)

\bibitem{BGP76}
Brochard, F., De~Gennes, P.G., Pfeuty, P.: Surface tension and deformations of
  membrane structures: relation to two-dimensional phase transitions.
\newblock J. de Physique \textbf{37}, 1099 (1976)

\bibitem{Brochard75}
Brochard, F., Lennon, J.F.: Frequency spectrum of flicker phenomenon in
  erythrocytes.
\newblock {J}. de {P}hysique \textbf{36}(11), 1035--1047 (1975)

\bibitem{BrownLondon98-2}
Brown, D.A., London, E.: Functions of lipid rafts in biological membranes.
\newblock Ann. Rev. Cell Develop. Biol. \textbf{14}, 111--136 (1998)

\bibitem{BrownLondon98-1}
Brown, D.A., London, E.: {Structure and origin of ordered lipid domains in
  biological membranes}.
\newblock {J. Mem. Biol.} \textbf{{164}}({2}), {103--114} ({1998})

\bibitem{BrownLondon00}
Brown, D.A., London, E.: Structure and function of sphingolipid- and
  cholesterol-rich membrane rafts.
\newblock J. Biol. Chem. \textbf{275}(23), 17,221--17,224 (2000)

\bibitem{Brown-elastic-review08}
Brown, F.L.H.: {Elastic Modeling of biomembranes and lipid bilayers}.
\newblock {Ann Rev. Phys. Chem.} \textbf{{59}}, {685--712} ({2008})

\bibitem{NA00}
C.~Nielsen, O.S.A.: Inclusion-induced bilayer deformations: Effects of
  monolayer equilibrium curvature.
\newblock Biophys. J. \textbf{79}, 2583--2604 (2000)

\bibitem{NGA98}
C.~Nielsen M.~Goulian, O.S.A.: Energetics of inclusion-induced bilayer
  deformations.
\newblock Biophys. J. \textbf{74}, 1966--1983 (2000)

\bibitem{CLN94}
Cai, W., Lubensky, T.C., Nelson, P., Powers., T.: Measure factors, tension, and
  correlations of fluid membranes.
\newblock J. de Physique II \textbf{4}, 931--949 (1994)

\bibitem{Canham70}
Canham, P.B.: Elastic properties of lipid bilayers---theory and possible
  experiments.
\newblock {J}. {T}heoret. {B}iol. \textbf{26}(1), 61--81 (1970)

\bibitem{Cantor97}
Cantor, R.S.: Lateral pressures in cell membranes: A mechanism for modulation
  of protein function.
\newblock J. Phys. Chem. B \textbf{101}, 1723--1725 (1997)

\bibitem{CG-02}
Capovilla, R., Guven, J.: Stresses in lipid membranes.
\newblock J. Phys. A: Math. Gen. \textbf{35}, 6233--6247 (2002)

\bibitem{CG-04}
Capovilla, R., Guven, J.: Stress and geometry of lipid vesicles.
\newblock J. Phys.: Condens. Matter \textbf{16}, S2187--S2191 (2004)

\bibitem{doCarmo}
do~Carmo, M.: Differential Geometry of Curves and Surfaces.
\newblock Prentice Hall, Englewood Cliffs, NJ (1976)

\bibitem{CPM11}
Chen, R., Poger, D., Mark, A.E.: Effect of high pressure on fully hydrated dppc
  and popc bilayers.
\newblock J. Phys. Chem. B \textbf{115}, 1038--1044 (2011)

\bibitem{CSS05}
Chiang, C., Shirinian, L., Sukharev, S.: Capping transmembrane helices of mscl
  with aromatic residues changes channel response to membrane stretch.
\newblock Biochemistry \textbf{44}, 12,589--12,597 (2005)

\bibitem{Chu05}
Chu, N., Ku\v{c}erka, N., Liu, Y.F., Tristram-Nagle, S., Nagle, J.F.: Anomalous
  swelling of lipid bilayer stacks is caused by softening of the bending
  modulus.
\newblock {P}hys. {R}ev. E \textbf{71}(4), 041,904 (2005)

\bibitem{Cooke05}
Cooke, I.R., Deserno, M.: Solvent-free model for self-assembling fluid bilayer
  membranes: Stabilization of the fluid phase based on broad attractive tail
  potentials.
\newblock {J}. {C}hem. {P}hys. \textbf{123}(22), 224,710 (2005)

\bibitem{Cooke06}
Cooke, I.R., Deserno, M.: Coupling between lipid shape and membrane curvature.
\newblock {B}iophys. {J}. \textbf{91}(2), 487--495 (2006)

\bibitem{Deserno2005}
Cooke, I.R., Kremer, K., Deserno, M.: Tunable generic model for fluid bilayer
  membranes.
\newblock {P}hys. {R}ev. E \textbf{72}(1), 011,506 (2005)

\bibitem{CP07}
Cordomi, A., Perez, J.J.: Molecular dynamics simulations of rhodopsin in
  different one-component lipid bilayers.
\newblock J. Phys. Chem. B \textbf{111}, 7052--7063 (2007)

\bibitem{Cuvelier05}
Cuvelier, D., Der\'{e}nyi, I., Bassereau, P., Nassoy, P.: Coalescence of
  membrane tethers: experiments, theory, and applications.
\newblock {B}iophys. {J}. \textbf{88}(4), 2714--2726 (2005)

\bibitem{CM11}
Cybulski, L.E., de~Mendoza, D.: Bilayer hydrophobic thickness and integral
  membrane protein function.
\newblock Current Protein \& Peptice Science \textbf{12}, 760--766 (2011)

\bibitem{Daleke-07}
Daleke, D.L.: Phospholipid flippases.
\newblock J. Biol. Chem. \textbf{282}(2), 821--825 (2007)

\bibitem{DBP94}
Dan, N., Berman, A., Pincus, P., Safran, S.A.: Membrane-induced interactions
  between inclusions.
\newblock J. de Physique II \textbf{4}, 1713 (1994)

\bibitem{DPS93}
Dan, N., Pincus, P., Safran, S.A.: Membrane-induced interactions between
  inclusions.
\newblock Langmuir \textbf{9}, 2768--2771 (1993)

\bibitem{MSC-III-09}
Das, A., Andersen, H.C.: {The multiscale coarse-graining method. III. A test of
  pairwise additivity of the coarse-grained potential and of new basis
  functions for the variational calculation}.
\newblock J. Chem. Phys. \textbf{131}(3), 034,102 (2009)

\bibitem{DL91}
David, F., Leibler, S.: Vanishing tension of fluctuating membranes.
\newblock J. de Physique II \textbf{1}, 959--976 (1991)

\bibitem{PK03}
De~Planque, M.R.R., Killian, J.A.: Protein-lipid interactions studied with
  designed transmembrane peptides: role of hydrophobic matching and interfacial
  anchoring (review).
\newblock Molecular Membrane Biology \textbf{20}, 271--284 (2003)

\bibitem{Debnath.submitted}
Debnath, A., Wiegand, S., Paulsen, H., Kremer, K., Peter, C.: Dual-scale
  atomistic and coarse grained simulations of chlorophyll/lipid systems.
\newblock submitted to J. Chem. Theory Comput.  (2013)

\bibitem{delgado-buscalioni08}
Delgado-Buscalioni, R., Kremer, K., Praprotnik, M.: Concurrent triple-scale
  simulation of molecular liquids.
\newblock J. Chem. Phys. \textbf{128}(11), 114,110 (2008)

\bibitem{delgado-buscalioni09}
Delgado-Buscalioni, R., Kremer, K., Praprotnik, M.: Coupling atomistic and
  continuum hydrodynamics through a mesoscopic model: Application to liquid
  water.
\newblock J. Chem. Phys. \textbf{131}(24), 244,107 (2009)

\bibitem{Deserno-wrapping-04}
Deserno, M.: Elastic deformation of a fluid membrane upon colloid binding.
\newblock Phys. Rev. E \textbf{69}, 031,903 (2004)

\bibitem{Deserno_BiomembraneFrontiers09}
Deserno, M.: Membrane elasticity and mediated interactions in continuum theory:
  A differential geometric approach.
\newblock In: R.~Faller, T.~Jue, M.L. Longo, S.H. Risbud (eds.) Biomembrane
  Frontiers: Nanostructures, Models, and the Design of Life, vol.~2, pp.
  41--74. Springer, New York (2009)

\bibitem{Deserno09}
Deserno, M.: Mesoscopic membrane physics: Concepts, simulations, and selected
  applications.
\newblock {M}acromol. {R}apid {C}omm. \textbf{30}(9-10), 752--771 (2009)

\bibitem{Deserno-etal07}
Deserno, M., M\"uller, M.M., Guven, J.: Contact lines for fluid surface
  adhesion.
\newblock Phys. Rev. E \textbf{76}, 011,605 (2007).
\newblock \doi{10.1103/PhysRevE.76.011605}

\bibitem{Deuling76}
Deuling, H., Helfrich, W.: The curvature elasticity of fluid membranes: A
  catalogue of vesicle shapes.
\newblock J. de Physique \textbf{37}, 1335--1345 (1976)

\bibitem{DEVA09.1}
DeVane, R., Shinoda, W., Moore, P.B., Klein, M.L.: {Transferable Coarse Grain
  Nonbonded Interaction Model for Amino Acids}.
\newblock J. Chem. Theory Comput. \textbf{5}, 2115--2124 (2009)

\bibitem{Diamant11}
Diamant, H.: Model-free thermodynamics of fluid vesicles.
\newblock Phys. Rev. E \textbf{84}(6), 0611,203 (2011)

\bibitem{Dockter.2012}
Dockter, C., Mueller, A.H., Dietz, C., Volkov, A., Polyhach, Y., Jeschke, G.,
  Paulsen, H.: Rigid core and flexible terminus: Structure of solubilized
  light-harvesting chlorophyll a/b complex ({LHCII}) measured by {EPR}.
\newblock J. Biol. Chem. \textbf{287}, 2915--2925 (2012).
\newblock \doi{10.1074/jbc.M111.307728}

\bibitem{Dockter.2009}
Dockter, C., Volkov, A., Bauer, C., Polyhach, Y., Joly-Lopez, Z., Jeschke, G.,
  Paulsen, H.: Refolding of the integral membrane protein light-harvesting
  complex ii monitored by pulse epr.
\newblock Proc. Natl. Acad. Sci. \textbf{106}(44), 18,485--18,490 (2009)

\bibitem{Dolezel10}
Dolezel, S.: Computer simulation of lipid bilayers.
\newblock Diploma thesis, Universit\"at Mainz (2010)

\bibitem{DommersnesFournier02}
{Dommersnes}, P., {Fournier}, J.: {The Many-Body Problem for Anisotropic
  Membrane Inclusions and the Self-Assembly of ``Saddle'' Defects into an ``Egg
  Carton''}.
\newblock Biophys. J. \textbf{83}, 2898--2905 (2002)

\bibitem{DommersnesFournier99}
Dommersnes, P., Fournier, J.B.: N-body study of anisotropic membrane
  inclusions: Membrane mediated interactions and ordered aggregation.
\newblock Eur. Phys. J. E \textbf{12}, 9--12 (1999)

\bibitem{DomFouEPL}
Dommersnes, P.G., Fournier, J.B.: Casimir and mean-field interactions between
  membrane inclusions subject to external torques.
\newblock Europhys. Lett. \textbf{46}, 256 (1999)

\bibitem{DS01}
D\"uchs, D., Schmid, F.: Phase behavior of amphiphilic monolayers: Theory and
  simulations.
\newblock J. Phys.: Cond. Matt. \textbf{13}, 4853 (2001)

\bibitem{DLT99}
Dumas, F., Lebrun, M.C., Tocanne, J.F.: Is the protein/lipid hydrophobic
  matching principle relevant to membrane organization and functions?
\newblock FEBS Letters \textbf{458}, 271--277 (1999)

\bibitem{END83}
Elliott, J., Needham, D., Dilger, J., Haydon, D.: The effects of bilayer
  thickness and tension on gramicidin single-channel lifetimes.
\newblock Biochimica et Biophysica Acta \textbf{735}, 95--103 (1983)

\bibitem{Evans90}
Evans, E., Rawicz, W.: Entropy-driven tension and bending elasticity in
  condensed-fluid membranes.
\newblock {P}hys. {R}ev. {L}ett. \textbf{64}(17), 2094--2097 (1990)

\bibitem{Evans74}
Evans, E.A.: Bending resistance and chemically-induced moments in membrane
  bilayers.
\newblock {B}iophys. {J}. \textbf{14}(12), 923--931 (1974)

\bibitem{Farago03}
Farago, O.: {``}water-free{''} computer model for fluid bilayer membranes.
\newblock {J}. {C}hem. {P}hys. \textbf{119}(1), 596--605 (2003)

\bibitem{Farago11}
Farago, O.: Mechanical surface tension governs membrane thermal fluctuations.
\newblock Phys. Rev. E \textbf{84}, 051,944 (2011)

\bibitem{FP03}
Farago, O., Pincus, P.: The effect of thermal fluctuations on schulman area
  elasticity.
\newblock Eur. Phys. J. E \textbf{11}, 399--408 (2003)

\bibitem{FP04}
Farago, O., Pincus, P.: Statistical mechanics of bilayer membrane with a fixed
  projected area.
\newblock J. Chem. Phys. \textbf{120}, 2934--2950 (2004)

\bibitem{FB93}
Fattal, D.R., Ben-Shaul, A.: A molecular model for lipid-protein interaction in
  membranes: the role of hydrophobic mismatch.
\newblock Biophys. J. \textbf{65}, 1795--1809 (1993)

\bibitem{FBS95}
Fattal, D.R., Ben-Shaul, A.: Lipid chain packing and lipid-protein interaction
  in membranes.
\newblock Physica A \textbf{220}, 192--216 (1995)

\bibitem{Faucon89}
Faucon, J.F., Mitov, M.D., M\'{e}l\'{e}ard, P., Bivas, I., Bothorel, P.:
  Bending elasticity and thermal fluctuations of lipid-membranes---theoretical
  and experimental requirements.
\newblock {J}. de {P}hysique \textbf{50}(17), 2389--2414 (1989)

\bibitem{Foerster86}
F\"orster, D.: {On the scale dependence, due to thermal fluctuations, of the
  elastic properties of membranes}.
\newblock Phys. Lett. A \textbf{114}(3), 115--120 (1986)

\bibitem{Fournier98}
Fournier, J.B.: Coupling between membrane tilt-difference and dilation: A new
  ``ripple'' instability and multiple crystalline inclusions phases.
\newblock EPL \textbf{43}, 725--730 (1998)

\bibitem{Fournier99}
Fournier, J.B.: Microscopic membrane elasticity and interactions among membrane
  inclusions: Interplay between the shape, dilation, tilt and tilt-difference
  modes.
\newblock Eur. Phys. J. B \textbf{11}, 261--272 (1999)

\bibitem{Fournier12}
Fournier, J.B.: Comment on "are stress-free membranes really 'tensionless'?" by
  {S}chmid {F}.
\newblock EPL \textbf{97}, 18,001 (2012)

\bibitem{FB08}
Fournier, J.B., C., B.: Direct calculation from the stress tensor of the
  lateral surface tension of fluctuating fluid membranes.
\newblock Phys. Rev. Lett. \textbf{100}, 078,103 (2008)

\bibitem{FournierDommersnes97}
Fournier, J.B., Dommersnes, P.G.: Comment on ``long-range forces in
  heterogeneous fluid membranes''.
\newblock Europhys. Lett. \textbf{39}(6), 681 (1997)

\bibitem{Fournier-etal03}
Fournier, J.B., Dommersnes, P.G., P., G.: Dynamin recruitment by clathrin
  coats: a physical step?
\newblock C. R. Biol. \textbf{326}, 467--476 (2003)

\bibitem{Fromherz83}
Fromherz, P.: Lipid-vesicle structure: size control by edge-active agents.
\newblock Chem. Phys. Lett. \textbf{94}, 259--266 (1983)

\bibitem{Galley-etal-10}
Galley, C.R., Leibovich, A.K., Rothstein, I.Z.: Finite size corrections to the
  radiation reaction force in classical electrodynamics.
\newblock Phys. Rev. Lett. \textbf{105}, 094,802 (2010)

\bibitem{Garab.2000}
Garab, G., Lohner, K., Laggner, P., Farkas, T.: Self-regulation of the lipid
  content of membranes by non-bilayer lipids: a hypothesis.
\newblock Trends Plant Sci. \textbf{5}(11), 489--494 (2000)

\bibitem{Gebhardt-etal-77}
Gebhardt, C., Gruler, H., Sackmann, E.: {Domain-Structure and Local Curvature
  in Lipid Bilayers and Biological Membranes}.
\newblock {Z. Naturforsch. C} \textbf{{32}}({7-8}), {581--596} ({1977})

\bibitem{Genco93}
Genco, I., Gliozzi, A., Relini, A., Robello, M., Scalas, E.: Osmotic-pressure
  induced pores in phospholipid vesicles.
\newblock {B}iochim. et {B}iophys. {A}cta \textbf{1149}(1), 10--18 (1993)

\bibitem{Gil-etal-98}
Gil, T., Ipsen, J.H., Mouritsen, O.G., Sabra, M.C., Sperotto, M.M., Zuckermann,
  M.J.: Theoretical analysis of protein organization in lipid membranes.
\newblock Biochim. Biophys. Acta \textbf{1376}(3), 245--266 (1998)

\bibitem{Gil-etal-97}
Gil, T., Sabra, M.C., Ipsen, J.H., Mouritsen, O.G.: Wetting and capillary
  condensation as means of protein organization in membranes.
\newblock Biophys, J. \textbf{73}(4), 1728--1741 (1997)

\bibitem{GGL99}
Goetz, R., Gompper, G., Lipowsky, R.: Mobility and elasticity of self-assembled
  membranes.
\newblock Phys. Rev. Lett. \textbf{82}, 221--224 (1999)

\bibitem{GoldbergerRothstein06}
Goldberger, W.D., Rothstein, I.Z.: An effective field theory of gravity for
  extended objects.
\newblock Phys. Rev. D \textbf{73}, 104,029 (2006)

\bibitem{GompperKlein92}
Gompper, G., Klein, S.: Ginzburg-landau theory of aqueous surfactant solutions.
\newblock {J}. {P}hys. {II} (France) \textbf{2}, 1725--1744 (1992)

\bibitem{GompperKroll98}
Gompper, G., Kroll, D.: {Membranes with fluctuating topology: Monte Carlo
  simulations}.
\newblock {Phys. Rev. Lett.} \textbf{{81}}({11}), {2284--2287} ({1998})

\bibitem{GompperKroll97}
Gompper, G., Kroll, D.M.: {Network models of fluid, hexatic and polymerized
  membranes}.
\newblock {J. Phys.: Condens. Matt.} \textbf{{9}}({42}), {8795--8834} ({1997})

\bibitem{Gosselin2012}
Gosselin, H.P., Morbach, H., M\"uller, M.M.: Interface-mediated interactions:
  Entropic forces of curved membranes.
\newblock Phys. Rev. E \textbf{83}, 051,921 (2012)

\bibitem{GouBruPinEPL}
Goulian, M., Bruinsma, R., Pincus, P.: Long-range forces in heterogeneous fluid
  membranes.
\newblock Europhys. Lett. \textbf{22}, 145 (1993)

\bibitem{GouBruPinEPLe}
Goulian, M., Bruinsma, R., Pincus, P.: Long-range forces in heterogeneous fluid
  membranes: Erratum.
\newblock Europhys. Lett. \textbf{23}, 155 (1993)

\bibitem{Guven-aux-04}
Guven, J.: Membrane geometry with auxiliary variables and quadratic
  constraints.
\newblock J. Phys. A: Math. Gen. \textbf{37}, L313--L319 (2004)

\bibitem{Hancock06}
Hancock, J.F.: Lipid rafts: Contentious only from simplistic standpoints.
\newblock Nat. Rev. Mol. Cell Biol. \textbf{7}, 456--462 (2006)

\bibitem{Harmandaris06}
Harmandaris, V.A., Deserno, M.: A novel method for measuring the bending
  rigidity of model lipid membranes by simulating tethers.
\newblock {J}. {C}hem. {P}hys. \textbf{125}(20), 204,905 (2006)

\bibitem{HHW99}
Harroun, T.A., Heller, W.T., Weiss, T.M., Yang, L., Huang, H.W.: Experimental
  evidence for hydrophobic matching and membrane-mediated interactions in lipid
  bilayers containing gramicidin.
\newblock Biophys. J. \textbf{76}, 937--945 (1999)

\bibitem{HHW99b}
Harroun, T.A., Heller, W.T., Weiss, T.M., Yang, L., Huang, H.W.: Theoretical
  analysis of hydrophobic matching and membrane-mediated interactions in lipid
  bilayers containing gramicidin.
\newblock Biophys. J. \textbf{76}, 3176--3185 (1999)

\bibitem{HB00}
Harzer, U., Bechinger, B.: Alignment of lysine-anchored membrane peptides under
  conditions of hydrophobic mismatch: A cd, $^{15}$n and $^{31}$p solid-state
  nmr spectroscopy investigation.
\newblock Biochemistry \textbf{39}, 13,106--13,114 (2000)

\bibitem{Helfrich73}
Helfrich, W.: Elastic properties of lipid bilayers---theory and possible
  experiments.
\newblock {Z}. {N}aturforsch. C \textbf{28}(11), 693--703 (1973)

\bibitem{Helfrich74}
Helfrich, W.: The size of bilayer vesicles generated by sonication.
\newblock {P}hys. {L}ett. A \textbf{50}(2), 115--116 (1974)

\bibitem{Helfrich81}
Helfrich, W.: Physics of Defects.
\newblock North Holland, {A}msterdam (1981)

\bibitem{Helfrich94}
Helfrich, W.: Lyotropic lamellar phases.
\newblock {J}. {P}hys. {C}ondens. {M}att. \textbf{6}, A79--A92 (1994)

\bibitem{HelfrichWeikl01}
Helfrich, W., Weikl, T.R.: Two direct methods to calculate fluctuation forces
  between rigid objects embedded in fluid membranes.
\newblock Eur. Phys. J. E \textbf{5}, 423--439 (2001)

\bibitem{Helfrich85}
{Helfrich, W.}: Effect of thermal undulations on the rigidity of fluid
  membranes and interfaces.
\newblock J. Phys. France \textbf{46}(7), 1263--1268 (1985)

\bibitem{Henriksen04}
Henriksen, J., Rowat, A.C., Ipsen, J.H.: Vesicle fluctuation analysis of the
  effects of sterols on membrane bending rigidity.
\newblock {E}ur. {B}iophys. {J}. \textbf{33}(8), 732--741 (2004)

\bibitem{HK10}
Holt, A., Killian, J.A.: Orientation and dynamics of transmembrane peptides:
  The power of simple models.
\newblock European Biophysical Journal \textbf{39}, 609--621 (2010)

\bibitem{Horn.2007}
Horn, R., Grundmann, G., Paulsen, H.: Consecutive binding of chlorophylls a and
  b during the assembly in vitro of light-harvesting chlorophyll-a/b protein
  (lhciib).
\newblock J. Mol. Biol. \textbf{366}, 1045--1054 (2007)

\bibitem{Horn.2004}
Horn, R., Paulsen, H.: Early steps in the assembly of light-harvesting
  chlorophyll a/b complex.
\newblock J. Biol. Chem. \textbf{279}(43), 44,400--44,406 (2004)

\bibitem{HuBriguglioDeserno12}
Hu, M., Briguglio, J.J., Deserno, M.: Determining the gaussian curvature
  modulus of lipid membranes in simulations.
\newblock {B}iophys. {J}. \textbf{102}(6), 1403--1410 (2012)

\bibitem{HuDigginsDeserno13}
Hu, M., Diggins~IV, P., Deserno, M.: Determining the bending modulus of a lipid
  membrane by simulating buckling.
\newblock J. Chem. Phys. \textbf{138}, ??? (2013).
\newblock \doi{10.1063/1.4808077}

\bibitem{HuDeJongMarrinkDeserno12}
Hu, M., de~Jong, D.H., Marrink, S.J., Deserno, M.: Gaussian curvature
  elasticity determined from global shape transformations and local stress
  distributions: A comparative study using the martini model.
\newblock Farad. Discuss. \textbf{161}, 365--382 (2013)

\bibitem{H86}
Huang, H.W.: Deformation free energy of bilayer membrane and its effect on
  gramicidin channel lifetime.
\newblock Biophys. J. \textbf{50}, 1061--1070 (1986)

\bibitem{H95}
Huang, H.W.: Elasticity of lipid bilayer interacting with amphiphilic helical
  peptides.
\newblock J. de Physique II \textbf{5}, 1427--1431 (1995)

\bibitem{Imparato06}
Imparato, A.: Surface tension in bilayer membranes with fixed projected area.
\newblock J. Chem. Phys. \textbf{124}, 154,714 (2006)

\bibitem{IKM87}
Ipsen, J.H., Karlstr\"om, G., Mouritsen, O.B., Wennerstr\"om, H., Zuckermann,
  M.J.: Phase equilibria in the phosphatidylcholine-cholesterol system.
\newblock Biochimica et Biophysica Acta \textbf{905}, 162--172 (1987)

\bibitem{IMN76}
Israelachvili, J.N., Mitchell, D.J., Ninham, B.W.: Theory of self-assembly of
  hydrocarbon amphiphiles into micelles and bilayers.
\newblock J. Chem. Soc.{,} Faraday Trans. 2 \textbf{72}, 1525--1568 (1976)

\bibitem{IzvekovVoth05}
Izvekov, S., Both, G.A.: Multiscale coarse graining of liquid-state systems.
\newblock J. Chem. Phys. \textbf{123}(13), 134,105 (2005)

\bibitem{IZVE10.1}
Izvekov, S., Chung, P.W., Rice, B.M.: {The multiscale coarse-graining method:
  Assessing its accuracy and introducing density dependent coarse-grain
  potentials}.
\newblock J. Chem. Phys. \textbf{133}, 064,109 (2010)

\bibitem{Jaehnig96}
J{\"a}hnig, F.: What is the surface tension of a lipid bilayer membrane?
\newblock Biophys. J. \textbf{71}, 1348--1349 (1996)

\bibitem{JKN10}
Jamroz, D., Kepczynski, M., Nowakowska, M.: Molecular structure of the
  dioctadecyldimethylammonium bromide (dodab) bilayer.
\newblock Langmuir \textbf{26}, 15,076--15,079 (2010)

\bibitem{JM04}
Jensen, M.O., Mouritsen, O.G.: Lipids do influence protein function -- the
  hydrophobic matching hypothesis revisited.
\newblock Biochimica et Biophysica Acta \textbf{1666}, 205--226 (2004)

\bibitem{Kindt04}
Jiang, F.Y., Bouret, Y., Kindt, J.T.: Molecular dynamics simulations of the
  lipid bilayer edge.
\newblock {B}iophys. {J}. \textbf{87}(1), 182--192 (2004)

\bibitem{John07.1}
Johnson, M.E., Head-Gordon, T., Louis, A.A.: Representability problems for
  coarse-grained water potentials.
\newblock J. Chem. Phys. \textbf{126}(14), 144,509 (2007)

\bibitem{vanKampen}
van Kampen, N.G.: Stochastic Processes in Physics and Chemistry, 3 edn.
\newblock Elsevier, Amsterdam (2007)

\bibitem{Karatekin03}
Karatekin, E., Sandre, O., Guitouni, H., Borghi, N., Puech, P.H.,
  Brochard-Wyart, F.: Cascades of transient pores in giant vesicles: Line
  tension and transport.
\newblock {B}iophys. {J}. \textbf{84}(3), 1734--1749 (2003)

\bibitem{GolestanianKardar99}
Kardar, M., Golestanian, R.: The ``friction'' of vacuum, and other
  fluctuation-induced forces.
\newblock Rev. Mod. Phys. \textbf{71}, 1233--1245 (1999)

\bibitem{KTL00}
Katsaras, J., Tristram-Nagle, S., Liu, Y., Headrick, R.L., Fontes, E., Mason,
  P.C., Nagle, J.F.: Clarification of the ripple phase of lecithin bilayers
  using fully hydrated, aligned samples.
\newblock Phys. Rev. E \textbf{61}, 5668 (2000)

\bibitem{KI98}
Killian, J.A.: Hydrophobic mismatch between proteins and lipids in membranes.
\newblock Biochimica et Biophysica Acta \textbf{1376}, 401--416 (1998)

\bibitem{KimChouRudnick08}
Kim, K.S., Chou, T., Rudnick, J.: Degenerate ground-state lattices of membrane
  inclusions.
\newblock Phys. Rev. E \textbf{78}, 011,401 (2008).
\newblock \doi{10.1103/PhysRevE.78.011401}

\bibitem{Kim98}
Kim, K.S., Neu, J., Oster, G.: Curvature-mediated interactions between membrane
  proteins.
\newblock Biophys. J. \textbf{75}(5), 2274--2291 (1998)

\bibitem{KimNeuOster99}
Kim, K.S., Neu, J.C., Oster, G.F.: Many-body forces between membrane
  inclusions: A new pattern-formation mechanism.
\newblock Europhys. Lett. \textbf{48}(1), 99--105 (1999)

\bibitem{Kleinert1986}
Kleinert, H.: Thermal softening of curvature elasticity in membranes.
\newblock Physs Lett. A \textbf{114}(5), 263--268 (1986)

\bibitem{KoltoverRaedlerSafinya99}
Koltover, I., R\"{a}dler, J.O., Safinya, C.R.: Membrane mediated attraction and
  ordered aggregation of colloidal particles bound to giant phospholipid
  vesicles.
\newblock Phys. Rev. Lett. \textbf{82}, 1991--1994 (1999)

\bibitem{KC94}
Koynova, R., Caffrey, M.: Phases and phase transitions of the hydrated
  phosphatidylethanolamines.
\newblock Chemistry and Physics of Lipids \textbf{69}, 1--34 (1994)

\bibitem{KC98}
Koynova, R., Caffrey, M.: Phases and phase transitions of the
  phosphatidylcholines.
\newblock Biochimica et Biophysica Acta - Reviews on Biomembranes
  \textbf{1376}, 91--145 (1998)

\bibitem{Kozlovsky-etal-04}
Kozlovsky, Y., Zimmerberg, J., Kozlov, M.M.: {Orientation and interaction of
  oblique cylindrical inclusions embedded in a lipid monolayer: A theoretical
  model for viral fusion peptides}.
\newblock {Biophys. J.} \textbf{{87}}({2}), {999--1012} ({2004})

\bibitem{Kramer71}
Kramer, L.: Theory of light scattering from fluctuations of membranes and
  monolayers.
\newblock J. Chem. Phys. \textbf{55}, 2097--2105 (1971)

\bibitem{KS05}
Kranenburg, M., Smit, B.: Phase behavior of model lipid bilayers.
\newblock J. Phys. Chem. B \textbf{109}, 6553--6563 (2005)

\bibitem{Kreyszig}
Kreyszig, E.: Differential Geometry.
\newblock Dover, New York (1991)

\bibitem{KumarGompperLipowsky01}
Kumar, P.B.S., Gompper, G., Lipowsky, R.: {Budding dynamics of multicomponent
  membranes}.
\newblock {Phys. Rev. Lett.} \textbf{{86}}({17}), {3911--3914} ({2001})

\bibitem{DR92}
Le~Doussal, P., Radzihovsky, L.: {S}elf-consistent theory of polymerized
  membranes.
\newblock Phys. Rev. Lett. \textbf{69}(8), 1209--1211 (1992)

\bibitem{Lee05}
Lee, A.G.: How lipids and proteins interact in a membrane: A molecular
  approach.
\newblock Mol. Biosyst. \textbf{3}, 203--212 (2005)

\bibitem{spiderman}
Lee, S.: Spiderman, Amazing Fantasy \# 15.
\newblock Marvel Comics, New York (1962)

\bibitem{LS05}
Lenz, O., Schmid, F.: A simple computer model for liquid lipid bilayers.
\newblock J. Mol. Liqu. \textbf{117}(1-3), 147--152 (2005)

\bibitem{LS07}
Lenz, O., Schmid, F.: Structure of symmetric and asymmetric ripple phases in
  lipid bilayers.
\newblock Phys. Rev. Lett. \textbf{98}, 058,104 (2007)

\bibitem{Leslie11}
Leslie, M.: Do lipid rafts exist?
\newblock Science \textbf{334}, 1046--1047 (2011)

\bibitem{LiPastor2013}
Li, J., Pastor, K.A., Shi, A.C., Schmid, F., Zhou, J.: Elastic properties and
  line tension of self-assembled bilayer membranes.
\newblock preprint  (2013)

\bibitem{Lindahl00}
Lindahl, E., Edholm, O.: Mesoscopic undulations and thickness fluctuations in
  lipid bilayers from molecular dynamics simulations.
\newblock {B}iophys. {J}. \textbf{79}(1), 426--433 (2000)

\bibitem{LG93}
Lipowsky, R., Grotehans, S.: Hydration versus protrusion forces between lipid
  bilayers.
\newblock EPL \textbf{23}, 599--604 (1993)

\bibitem{LG93b}
Lipowsky, R., Grotehans, S.: Renormalization of hydration forces by collective
  protrusion modes.
\newblock Biophysical Chemistry \textbf{49}, 27--37 (1993)

\bibitem{Liu04}
Liu, Y.F., Nagle, J.F.: Diffuse scattering provides material parameters and
  electron density profiles of biomembranes.
\newblock {P}hys. {R}ev. E \textbf{69}(4), 040,901 (2004)

\bibitem{Liu.2004}
Liu, Z., Yan, H., Wang, K., Kuang, T., Zhang, J., Gul, L., An, X., Chang, W.:
  Crystal structure of spinach major light-harvesting complex at 2.72 agstrom
  resolution.
\newblock Nature \textbf{428}, 287--292 (2004)

\bibitem{LMK03}
Loison, C., Mareschal, M., Kremer, K., Schmid, F.: Thermal fluctuations in a
  lamellar phase of a binary amphiphile-solvent mixture: A molecular-dynamics
  study.
\newblock J. Chem. Phys. \textbf{119}, 13,138--13,148 (2003)

\bibitem{MARTINI-carbohydrates09}
L\'opez, C.A., Rzepiela, A.J., de~Vries, A.H., Dijkhuizen, L., Huenenberger,
  P.H., Marrink, S.J.: {Martini Coarse-Grained Force Field: Extension to
  Carbohydrates}.
\newblock {J. Chem. Theory Comput.} \textbf{{5}}({12}), {3195--3210} ({2009})

\bibitem{MARTINI-glyco13}
L\'opez, e.A., Sovova, Z., van Eerden, F.J., de~Vries, A.H., Marrink, S.J.:
  Martini force field parameters for glycolipids.
\newblock J. Chem. Theory Comput. \textbf{9}(3), 1694--1708 (2013)

\bibitem{MSC-LuVoth-12}
Lu, L., Voth, G.A.: The Multiscale Coarse-Graining Method, pp. 47--81.
\newblock John Wiley \&\ Sons, Inc. (2012)

\bibitem{LYUB10.1}
Lyubartsev, A., Mirzoev, A., Chen, L.J., Laaksonen, A.: {Systematic
  coarse-graining of molecular models by the Newton inversion method}.
\newblock Faraday Discuss \textbf{144}, 43--56 (2010)

\bibitem{LYUB95.1}
Lyubartsev, A.P., Laaksonen, A.: Calculation of effective interaction
  potentials from radial-distribution functions - a reverse {Monte-Carlo}
  approach.
\newblock Phys. Rev. E \textbf{52}, 3730 -- 3737 (1995)

\bibitem{Machta-etal-12}
Machta, B.B., Veatch, S.L., Sethna, J.P.: Critical casimir forces in cellular
  membranes.
\newblock Phys. Rev. Lett. \textbf{109}, 138,101 (2012)

\bibitem{Markin-81}
Markin, V.S.: {Lateral Organization of Membranes and Cell Shapes}.
\newblock {Biophys. J.} \textbf{{36}}({1}), {1--19} ({1981})

\bibitem{MarrinkVriesTieleman2009}
Marrink Siewert J. de~Vries, A.H., Tieleman, D.P.: Lipids on the move:
  Simulations of membrane pores, domains, stalks and curves.
\newblock Biochim. Biophys. Acta -- Biomembranes \textbf{1788}(1), 149--168
  (2009)

\bibitem{MM01}
Marrink, S.J., Mark, A.E.: Effect of undulations on surface tension in
  simulated bilayers.
\newblock J. Phys. Chem. B \textbf{105}, 6122--6127 (2001)

\bibitem{marrink_martini_2007}
Marrink, S.J., Risselada, H.J., Yefimov, S., Tieleman, D.P., de~Vries, A.H.:
  The {MARTINI} force field: Coarse grained model for biomolecular simulations.
\newblock {J}. {P}hys. {C}hem. B \textbf{111}, 7812--7824 (2007)

\bibitem{Marrink04}
Marrink, S.J., de~Vries, A.H., Mark, A.E.: Coarse grained model for
  semiquantitative lipid simulations.
\newblock {J}. {P}hys. {C}hem. B \textbf{108}(2), 750--760 (2004)

\bibitem{Marsh97}
Marsh, D.: Renormalization of the tension and area expansion modulus in fluid
  membranes.
\newblock Biophys. J. \textbf{73}, 865--869 (1997)

\bibitem{Marsh08b}
Marsh, D.: Energetics of hydrophobic matching in lipid-protein interactions.
\newblock Biophys. J. \textbf{94}, 3996--4013 (2008)

\bibitem{Marsh08}
Marsh, D.: Protein modulation of lipids, and vice-versa, in membranes.
\newblock Biochimica et Biophysica Acta \textbf{1778}, 1545--1575 (2008)

\bibitem{MNK07}
May, E.R., Narang, A., Kopelevich, D.I.: Role of molecular tilt in thermal
  fluctuations of lipid membranes.
\newblock Phys. Rev. E \textbf{76}, 021,913 (2007)

\bibitem{May-tilt-00}
May, S.: Protein-induced bilayer deformations: the lipid tilt degree of
  freedom.
\newblock Eur. Biophys. J. \textbf{29}(1), 17--28 (2000)

\bibitem{May00}
May, S.: Theories on structural perturbations of lipid bilayers.
\newblock Current Opinion in Colloid \& Interface Science \textbf{5}, 244--249
  (2000)

\bibitem{MayBenShaul-99}
May, S., Ben-Shaul, A.: Molecular theory of lipid-protein interaction and the
  lα-hii transition.
\newblock Biophys. J. \textbf{76}(2), 751--767 (1999)

\bibitem{MEGA11.1}
Megariotis, G., Vyrkou, A., Leygue, A., Theodorou, D.N.: {Systematic Coarse
  Graining of 4-Cyano-4 '-pentylbiphenyl}.
\newblock Ind. Eng. Chem. Res. \textbf{50}, 546--556 (2011)

\bibitem{MVS13}
Meinhardt, S., Vink, R., Schmid, F.: Monolayer curvature stabilizes nanoscale
  raft domains in mixed lipid bilayers.
\newblock PNAS \textbf{12}, 4476--4481 (2013)

\bibitem{MVS08}
de~Meyer, F.J.M., Venturoli, M., Smit, B.: Molecular simulations of
  lipid-mediated protein-protein interactions.
\newblock Biophys. J. \textbf{95}, 1851--1865 (2008)

\bibitem{MOGN08.1}
Mognetti, B.M., Yelash, L., Virnau, P., Paul, W., Binder, K., Mueller, M.,
  Macdowell, L.G.: {Efficient prediction of thermodynamic properties of
  quadrupolar fluids from simulation of a coarse-grained model: The case of
  carbon dioxide}.
\newblock J. Chem. Phys. \textbf{128}, 104,501 (2008)

\bibitem{MARTINI-proteins08}
Monticelli, L., Kandasamy, S.K., Periole, X., Larson, R.G., Tieleman, D.P.,
  Marrink, S.J.: {The MARTINI coarse-grained force field: Extension to
  proteins}.
\newblock {J. Chem. Theory Comput.} \textbf{{4}}({5}), {819--834} ({2008})

\bibitem{MB84}
Mouritsen, O.G., Bloom, M.: Mattress model of lipid-protein interactions in
  membranes.
\newblock Biophys. J. \textbf{36}, 141--153 (1984)

\bibitem{MullerKatsovSchick06}
M\"ueller, M., Katsov, K., Schick, M.: {Biological and synthetic membranes:
  What can be learned from a coarse-grained description?}
\newblock {Phys. Rep.} \textbf{{434}}({5-6}), {113--176} ({2006})

\bibitem{MUKH12.1}
Mukherje, B., L., D.S., K., K., Peter, C.: {Derivation of a Coarse Grained
  model for Multiscale Simulation of Liquid Crystalline Phase Transitions}.
\newblock J. Phys. Chem B submitted  (2012)

\bibitem{MullerDeserno10}
M\"uller, M.M., Deserno, M.: Cell model approach to membrane mediated protein
  interactions.
\newblock Progr. Theor. Phys. Suppl. \textbf{184}, 351--363 (2010)

\bibitem{MDG-05a}
M\"uller, M.M., Deserno, M., Guven, J.: Geometry of surface-mediated
  interactions.
\newblock Europhys. Lett. \textbf{69}, 482--488 (2005)

\bibitem{MDG-05b}
M\"uller, M.M., Deserno, M., Guven, J.: Interface-mediated interactions between
  particles: A geometrical approach.
\newblock Phys. Rev. E \textbf{72}, 061,407 (2005)

\bibitem{MullerDeserno07}
M\"uller, M.M., Deserno, M., Guven, J.: Balancing torques in membrane-mediated
  interactions: Exact results and numerical illustrations.
\newblock Phys. Rev. E \textbf{76}, 011,921 (2007).
\newblock \doi{10.1103/PhysRevE.76.011921}

\bibitem{FMP-review02}
M\"uller-Plathe, F.: Coarse-graining in polymer simulation: From the atomistic
  to the mesoscopic scale and back.
\newblock Chem. Phys. Chem. \textbf{3}(9), 754--769 (2002)

\bibitem{MULL09.1}
Mullinax, J.W., Noid, W.G.: {Extended ensemble approach for deriving
  transferable coarse-grained potentials}.
\newblock J. Chem. Phys. \textbf{131}, 104,110 (2009)

\bibitem{Munro03}
Munro, S.: {Lipid rafts: Elusive or illusive?}
\newblock Cell \textbf{115}(4), 377--388 (2003)

\bibitem{MURT09.1}
Murtola, T., Karttunen, M., Vattulainen, I.: {Systematic coarse graining from
  structure using internal states: Application to phospholipid/cholesterol
  bilayer}.
\newblock J. Chem. Phys. \textbf{131}, 055,101 (2009)

\bibitem{NNWS12}
Neder, J., Nielaba, P., West, B., Schmid, F.: Interactions of membranes with
  coarse-grain proteins: A comparison.
\newblock New J. of Physics \textbf{14}, 125,017 (2012)

\bibitem{NWN10}
Neder, J., West, B., Nielaba, P., Schmid, F.: Coarse-grained simulations of
  membranes under tension.
\newblock J. Chem. Phys. \textbf{132}, 115,101 (2010)

\bibitem{NWN11}
Neder, J., West, B., Nielaba, P., Schmid, F.: Membrane-mediated protein-protein
  interaction: A monte carlo study.
\newblock Current Nanoscience \textbf{7}, 656--666 (2010)

\bibitem{NP87}
Nelson, D.R., Peliti, L.: {F}luctuations in membranes with crystalline and
  hexatic order.
\newblock J. de Physique \textbf{48}(1), 1085--1092 (1987).
\newblock \doi{10.1051/jphys:019870048070108500}

\bibitem{NIEL03.1}
Nielsen, S.O., Lopez, C.F., Srinivas, G., Klein, M.L.: {A coarse grain model
  for n-alkanes parameterized from surface tension data}.
\newblock J. Chem. Phys. \textbf{119}, 7043--7049 (2003)

\bibitem{NimelaHyvonenVattulainen09}
Niemel\"a, P.S., Hyvonen, M.T., Vattulainen, I.: {Atom-scale molecular
  interactions in lipid raft mixtures}.
\newblock {Biochim. Biophys. Acta} \textbf{{1788}}({1}), {122--135} ({2009})

\bibitem{Noguchi-review-09}
Noguchi, H.: {Membrane Simulation Models from Nanometer to Micrometer Scale}.
\newblock {J. Phys. Soc. Japan} \textbf{{78}}({4}), {041,007} ({2009})

\bibitem{Noguchi-buckle-2011}
Noguchi, H.: Anisotropic surface tension of buckled fluid membranes.
\newblock Phys. Rev. E \textbf{83}, 061,919 (2011)

\bibitem{NoguchiGompper04}
Noguchi, H., Gompper, G.: {Fluid vesicles with viscous membranes in shear
  flow}.
\newblock {Phys. Rev. Lett.} \textbf{{93}}({25}), {258,102} ({2004})

\bibitem{MSC-I-08}
Noid, W.G., Chu, J.W., Ayton, G.S., Krishna, V., Izvekov, S., Voth, G.A., Das,
  A., Andersen, H.C.: {The multiscale coarse-graining method. I. A rigorous
  bridge between atomistic and coarse-grained models}.
\newblock J. Chem. Phys. \textbf{128}(24), 244,114 (2008)

\bibitem{MSC-II-08}
Noid, W.G., Liu, P., Wang, Y., Chu, J.W., Ayton, G.S., Izvekov, S., Andersen,
  H.C., Voth, G.A.: {The multiscale coarse-graining method. II. Numerical
  implementation for coarse-grained molecular models}.
\newblock J. Chem. Phys. \textbf{128}(24), 244,115 (2008)

\bibitem{O79}
Owicki, J.C., McConnell, H.M.: Theory of protein-lipid and protein-protein
  interactions in bilayer membranes.
\newblock PNAS \textbf{76}, 4750--4754 (1979)

\bibitem{O78}
Owicki, J.C., Springgate, M.W., McConnell, H.M.: Theoretical study of
  protein-lipid interactions in bilayer membranes.
\newblock PNAS \textbf{75}, 1616--1619 (1978)

\bibitem{ORL05}
\"Ozdirekcan, S., Rijkers, D.T.S., Liskamp, R.M.J., Killian, J.A.: Influence of
  flanking residues on tilt and rotation angles of transmembrane peptides in
  lipid bilayers. a solid-state $^2$h nmr study.
\newblock Biochemistry \textbf{44}, 1004--1012 (2005)

\bibitem{Pan08}
Pan, J., Tristram-Nagle, S., Ku\v{c}erka, N., Nagle, J.F.: Temperature
  dependence of structure, bending rigidity, and bilayer interactions of
  dioleoylphosphatidylcholine bilayers.
\newblock {B}iophys. {J}. \textbf{94}(1), 117--124 (2008)

\bibitem{P96}
Park, J.M.: {R}enormalization of fluctuating tilted hexatic membranes.
\newblock Phys. Rev. E \textbf{56}(7), R47--R50 (1996)

\bibitem{PL95}
Park, J.M., Lubensky, T.C.: {T}opological defects on fluctuating surfaces:
  General properties and the kosterlitz-thouless transition.
\newblock Phys. Rev. E \textbf{53}(3), 2648--2664 (1995)

\bibitem{PO05}
Park, S.H., Opella, S.J.: Tilt angle of a trans-membrane helix is determined by
  hydrophobic mismatch.
\newblock J. Mol. Biology \textbf{350}, 310--318 (2005)

\bibitem{ParLub96}
{Park, Jeong-Man}, {Lubensky, T. C.}: Interactions between membrane inclusions
  on fluctuating membranes.
\newblock J. Phys. I France \textbf{6}(9), 1217--1235 (1996)

\bibitem{PelitiLeibler85}
Peliti, L., Leibler, S.: Effects of thermal fluctuations on systems with small
  surface tension.
\newblock Phys. Rev. Lett. \textbf{54}, 1690--1693 (1985)

\bibitem{PETE08.1}
Peter, C., Delle~Site, L., Kremer, K.: Classical simulations from the atomistic
  to the mesoscale: coarse graining an azobenzene liquid crystal.
\newblock Soft Matter \textbf{4}, 859--869 (2008)

\bibitem{PeterKremer09}
Peter, C., Kremer, K.: Multiscale simulation of soft matter systems - from the
  atomistic to the coarse-grained level and back.
\newblock Soft Matter \textbf{5}, 4357--4366 (2009)

\bibitem{Pfeiffer93}
Pfeiffer, W., K\"{o}nig, S., Legrand, J.F., Bayerl, T., Richter, D., Sackmann,
  E.: Neutron spin echo study of membrane undulations in lipid multibilayers.
\newblock {E}urophys. {L}ett. \textbf{23}(6), 457--462 (1993)

\bibitem{Pike06}
Pike, L.: Rafts defined: A report on the keystone symposium on lipid rafts and
  cell function.
\newblock J. Lipid Res. \textbf{47}, 1597--1598 (2006)

\bibitem{poblete10}
Poblete, S., Praprotnik, M., Kremer, K., Delle~Site, L.: Coupling different
  levels of resolution in molecular simulations.
\newblock J. Chem. Phys. \textbf{132}(11), 114,101 (2010)

\bibitem{Porto-etal-11}
Porto, R.A., Ross, A., Rothstein, I.Z.: Spin induced multipole moments for the
  gravitational wave flux from binary inspirals to third post-newtonian order.
\newblock J. Cosmol. Astropart. Phys. \textbf{3}, 009 (2011)

\bibitem{PKF00}
Pralle, A., Keller, P., Florin, E.L., Simons, K., H\"orber, J.K.H.:
  Sphingolipid-cholesterol rafts diffuse as small entities in the plasma
  membrane of mammalian cells.
\newblock J. Cell Biol. \textbf{148}, 997--1007 (2000)

\bibitem{praprotnik05}
Praprotnik, M., Delle~Site, L., Kremer, K.: Adaptive resolution
  molecular-dynamics simulation: Changing the degrees of freedom on the fly.
\newblock J. Chem. Phys. \textbf{123}(22), 224,106 (2005)

\bibitem{praprotnik07}
Praprotnik, M., Delle~Site, L., Kremer, K.: A macromolecule in a solvent:
  Adaptive resolution molecular dynamics simulation.
\newblock J. Chem. Phys. \textbf{126}(13), 134,902 (2007)

\bibitem{praprotnik08}
Praprotnik, M., Site, L.D., Kremer, K.: Multiscale simulation of soft matter:
  From scale bridging to adaptive resolution.
\newblock Ann. Rev. Phys. Chem. \textbf{59}(1), 545--571 (2008)

\bibitem{HVK09}
R., H.S.A., L., V.S., L., K.S.: An introduction to critical points for
  biophysicists: Observations of compositional heterogeneity in lipid
  membranes.
\newblock Biochim. Biophys. Acta \textbf{1788}, 53--63 (2009)

\bibitem{Rawicz-etal-00}
Rawicz, W., Olbrich, K., McIntosh, T., Needham, D., Evans, E.: Effect of chain
  length and unsaturation on elasticity of lipid bilayers.
\newblock Biophys. J. \textbf{79}, 328--339 (2000)

\bibitem{REIT03.1}
Reith, D., Putz, M., M\"uller-Plathe, F.: Deriving effective mesoscale
  potentials from atomistic simulations.
\newblock J. Comp. Chem. \textbf{24}, 1624 -- 1636 (2003)

\bibitem{Reynwar08}
Reynwar, B.J., Deserno, M.: Membrane composition-mediated protein-protein
  interactions.
\newblock {B}iointerphases \textbf{3}, FA117--FA125 (2008)

\bibitem{ReynwarDeserno11}
Reynwar, B.J., Deserno, M.: Membrane-mediated interactions between circular
  particles in the strongly curved regime.
\newblock Soft Matter \textbf{7}, 8567--8575 (2011)

\bibitem{Reynwar07}
Reynwar, B.J., Illya, G., Harmandaris, V.A., M\"uller, M.M., Kremer, K.,
  Deserno, M.: Aggregation and vesiculation of membrane proteins by
  curvature-mediated interactions.
\newblock {N}ature \textbf{447}, 461--464 (2007)

\bibitem{Rheinstaedter06}
Rheinst\"{a}dter, M.C., H\"{a}u{\ss}ler, W., Salditt, T.: Dispersion relation
  of lipid membrane shape fluctuations by neutron spin-echo spectrometry.
\newblock {P}jys. {R}ev. {L}ett. \textbf{97}(4), 048,103 (2006)

\bibitem{TASI}
Rothstein, I.Z.: Tasi lectures on effective field theories (2003).
\newblock ArXiv:hep-ph/0308266

\bibitem{RW}
Rowlinson, J.S., Widom, B.: Molecular Theory of Capillarity, 1 edn.
\newblock Dover, New York (2002)

\bibitem{VOTCA09}
R\"uhle, V., Junghans, C., Lukyanov, A., Kremer, K., Andrienko, D.: Versatile
  object-oriented toolkit for coarse-graining applications.
\newblock J. Chem. Theo. Comput. \textbf{5}(12), 3211--3223 (2009)

\bibitem{SaizBandyopadhyayKlein02}
Saiz, L., Bandyopadhyay, S., Klein, M.L.: Towards an understanding of complex
  biological membranes from atomistic molecular dynamics simulations.
\newblock Biosci. Rep. \textbf{22}(2), 151--173 (2002)

\bibitem{SaizKlein02}
Saiz, L., Klein, M.L.: Computer simulation studies of model biological
  membranes.
\newblock Acc. Chem. Res. \textbf{35}(6), 482--489 (2002)

\bibitem{ST91}
Sankaram, M.B., Thompson, T.E.: Cholesterol-induced fluid-phase immiscibility
  in membranes.
\newblock PNAS \textbf{88}, 8686--8690 (1991)

\bibitem{SAVE09.1}
Savelyev, A., Papoian, G.A.: {Molecular renormalization group coarse-graining
  of electrolyte solutions: application to aqueous {NaCl} and {KCl}}.
\newblock J. Phys. Chem. B \textbf{113}, 7785--7793 (2009)

\bibitem{Schmid-review-09}
Schmid, F.: {Toy amphiphiles on the computer: What can we learn from generic
  models?}
\newblock {Macromol. Rapid. Comm.} \textbf{{30}}, {741--751} ({2009})

\bibitem{Schmid11}
Schmid, F.: Are stress-free membranes really ''tensionless''?
\newblock EPL \textbf{95}, 28,008 (2011)

\bibitem{Schmid12}
Schmid, F.: Reply to {C}omment on ''are stress-free membranes really
  tensionless?''.
\newblock EPL \textbf{97}, 18,002 (2012)

\bibitem{Schmid13}
Schmid, F.: Fluctuations in lipid bilayers: Are they understood?
\newblock Biophysical Reviews and Letters \textbf{at press} (2013).
\newblock \doi{10.1142/S179304801230011}

\bibitem{SDL07}
Schmid, F., D\"uchs, D., Lenz, O., West, B.: A generic model for lipid
  monolayers, bilayers, and membranes.
\newblock Comp. Phys. Comm. \textbf{177}(1-2), 168 (2007)

\bibitem{Schmid.2008}
Schmid, V.H.: Light-harvesting complexes of vascular plants.
\newblock Cellular and Molecular Life Sciences \textbf{65}(22), 3619--3639
  (2008)

\bibitem{SGW08}
Schmidt, U., Guigas, G., Weiss, M.: Cluster formation of transmembrane proteins
  due to hydrophobic mismatching.
\newblock Phys. Rev. Lett. \textbf{101}, 128,104 (2008)

\bibitem{SW10}
Schmidt, U., Weiss, M.: Hydrophobic mismatch-induced clustering as a primer for
  protein sorting in the secretory pathway.
\newblock Biophysical Chemistry \textbf{151}, 34--38 (2010)

\bibitem{Schneider84a}
Schneider, M.B., Jenkins, J.T., Webb, W.W.: Thermal fluctuations of large
  cylindrical phospholipid-vesicles.
\newblock {B}iophys. {J}. \textbf{45}(5), 891--899 (1984)

\bibitem{Schneider84b}
Schneider, M.B., Jenkins, J.T., Webb, W.W.: Thermal fluctuations of large
  quasi-spherical bimolecular phospholipid-vesicles.
\newblock {B}iophys. {J}. \textbf{45}(9), 1457--1472 (1984)

\bibitem{Scott2002}
Scott, H.: Modeling the lipid component of membranes.
\newblock Curr. Opin. Struct. Biol. \textbf{12}(4), 495--502 (2002)

\bibitem{Seifert-93}
Seifert, U.: Curvature-induced lateral phase segregation in two-component
  vesicles.
\newblock Phys. Rev. Lett. \textbf{70}, 1335--1338 (1993)

\bibitem{SeifertLanger93}
Seifert, U., Langer, S.A.: Viscous modes of fluid bilayer membranes.
\newblock Europhys. Lett. \textbf{23}(1), 71--76 (1993)

\bibitem{Semrau08}
Semrau, S., Idema, T., Holtzer, L., Schmidt, T., Storm, C.: Accurate
  determination of elastic parameters for multicomponent membranes.
\newblock {P}hys. {R}ev. {L}ett. \textbf{100}(8), 088,101 (2008)

\bibitem{SRK03}
Sengupta, K., Raghunathan, V.A., Katsaras, J.: Structure of the ripple phase of
  phospholipid multibilayers.
\newblock Phys. Rev. E \textbf{68}(3), 031,710 (2003)

\bibitem{SHEL08.1}
Shell, M.S.: {The relative entropy is fundamental to multiscale and inverse
  thermodynamic problems}.
\newblock J. Chem. Phys. \textbf{129}, 144,108 (2008)

\bibitem{SHEN11.1}
Shen, J.W., Li, C., van~der Vegt, N.F.A., Peter, C.: {Transferability of Coarse
  Grained Potentials: Implicit Solvent Models for Hydrated Ions}.
\newblock J. Chem. Theory Comput. \textbf{7}, 1916--1927 (2011)

\bibitem{ShibaNoguchi11}
Shiba, H., Noguchi, H.: Estimation of the bending rigidity and spontaneous
  curvature of fluid membranes in simulations.
\newblock Phys. Rev. E \textbf{84}, 031,926 (2011)

\bibitem{Siegel06}
Siegel, D.P.: Determining the ratio of the gaussian curvature and bending
  elastic moduli of phospholipids from q$_{\rm ii}$ phase unit cell dimensions.
\newblock {B}iophys. {J}. \textbf{91}(2), 608--618 (2006)

\bibitem{Siegel08}
Siegel, D.P.: The gaussian curvature elastic energy of intermediates in
  membrane fusion.
\newblock {B}iophys. {J}. \textbf{95}(11), 5200--5215 (2008)

\bibitem{Siegel04}
Siegel, D.P., Kozlov, M.M.: The gaussian curvature elastic modulus of
  n-monomethylated dioleoylphosphatidylethanolamine: Relevance to membrane
  fusion and lipid phase behavior.
\newblock {B}iophys. {J}. \textbf{87}(1), 366--374 (2004)

\bibitem{SILB06.1}
Silbermann, J.R., Klapp, S.H.L., Schoen, M., Chennamsetty, N., Bock, H.,
  Gubbins, K.E.: {Mesoscale modeling of complex binary fluid mixtures: Towards
  an atomistic foundation of effective potentials}.
\newblock J. Chem. Phys. \textbf{124}, 074,105 (2006)

\bibitem{SI97}
Simons, K., Ikonen, E.: Functional rafts in cell membranes.
\newblock Nature \textbf{387}, 569--572 (1997)

\bibitem{SK04}
Simons, K., Vaz, W.L.C.: model systems, lipid rafts, and cell membranes.
\newblock Annu. Rev. Biophys. Biomol. Struct. \textbf{33}, 269--295 (2004)

\bibitem{SingerNicolson72}
Singer, S.J., Nicolson, G.K.: {Fluid Mosaic Model of Structure of
  Cell-Membranes}.
\newblock {Science} \textbf{{175}}({4023}), {720--731} ({1972})

\bibitem{SmithTanford72}
Smith, R., Tanford, C.: {The critical micelle concentration of
  {L}$_\alpha$-di\-pal\-mi\-to\-yl\-phos\-pha\-ti\-dyl\-cho\-line in water and
  water/methanol solutions}.
\newblock J. Mol. Biol. \textbf{67}(1), 75--83 (1972)

\bibitem{SM91}
Sperotto, M.M., Mouritsen, O.G.: Monte carlo simulation of lipid order
  parameter profiles near integral membrane proteins.
\newblock Biophys. J. \textbf{59}, 261--270 (1991)

\bibitem{Standfuss.2005}
Standfuss, R., van Scheltinga, A.C.T., Lamborghini, M., K{\"u}hlbrandt, W.:
  Mechanisms of photoprotection and nonphotochemical quenching in pea
  light-harvesting complex at 2.5 {\aa} resolution.
\newblock EMBO J. \textbf{24}, 919--928 (2005)

\bibitem{Stecki08}
Stecki, J.: Balance of forces in simulated bilayers.
\newblock J. Phys. Chem. B \textbf{112}(14), 4246--4252 (2008)

\bibitem{SEU12}
Strandberg, E., Esteban-Martin, S., Ulrich, A.S., Salgado, J.: Hydrophobic
  mismatch of mobile transmembrane helices: Merging theory and experiments.
\newblock Biochimica et Biophysica Acta \textbf{1818}, 1242--1249 (2012)

\bibitem{STW09}
Strandberg, E., Tremouilhac, P., Wadhwani, P., Ulrich, A.S.: Synergistic
  transmembrane insertion of the heterodimeric pgla/magainin 2 complex studied
  by solid-state nmr.
\newblock Biochimica et Biophysica Acta \textbf{1788}, 1667--1679 (2009)

\bibitem{SG08}
Sun, X., Gezelter, J.D.: Dipolar ordering in the ripple phases of
  molecular-scale models of lipid membranes.
\newblock J. Phys. Chem. B \textbf{112}, 1968--1975 (2008)

\bibitem{Szleifer90}
Szleifer, I., Kramer, D., Ben-Shaul, A., Gelbart, W.M., Safran, S.A.: Molecular
  theory of curvature elasticity in surfactant films.
\newblock {J}. {C}hem. {P}hys. \textbf{92}, 6800--6817 (1990)

\bibitem{Takeda99}
Takeda, T., Kawabata, Y., Seto, H., Komura, S., Ghosh, S.K., Nagao, M.,
  Okuhara, D.: Neutron spin-echo investigations of membrane undulations in
  complex fluids involving amphiphiles.
\newblock {J}. {P}hys. {C}hem. {S}olids \textbf{60}(8-9), 1375--1377 (1999)

\bibitem{Taupin75}
Taupin, C., Dvolaitzky, M., Sauterey, C.: Osmotic-pressure induced pores in
  phospholipid vesicles.
\newblock {B}iochem. \textbf{14}(21), 4771--4775 (1975)

\bibitem{Templer98}
Templer, R.H., Khoo, B.J., Seddon, J.M.: Gaussian curvature modulus of an
  amphiphile monolayer.
\newblock {L}angmuir \textbf{14}(26), 7427--7434 (1998)

\bibitem{Tian08}
Tian, A., Baumgart, T.: Sorting of lipids and proteins in membrane curvature
  gradients.
\newblock {B}iophys. {J}. \textbf{96}(7), 2676--2688 (2008)

\bibitem{Briels04a}
Tolpekina, T.V., den Otter, W.K., Briels, W.J.: Simulations of stable pores in
  membranes: System size dependence and line tension.
\newblock {J}. {C}hem. {P}hys. \textbf{121}(16), 8014--8020 (2004)

\bibitem{TNagle07}
Tristram-Nagle, S., Nagle, J.F.: Hiv-1 fusion peptide decreases bending energy
  and promotes curved fusion intermediates.
\newblock {B}iophys. {J}. \textbf{93}(6), 2048--2055 (2007)

\bibitem{TSCH98.1}
Tsch\"op, W., Kremer, K., Batoulis, J., Burger, T., Hahn, O.: Simulation of
  polymer melts. i. coarse-graining procedure for polycarbonates.
\newblock Acta Polym. \textbf{49}(2-3), 61 -- 74 (1998)

\bibitem{TSS05}
Turner, M.S., Sens, P., Socci, N.D.: Nonequilibrium raft-like membrane domains
  under continuous recycling.
\newblock Phys. Rev. Lett. \textbf{95}, 168,301 (2005)

\bibitem{VeatchKeller03}
Veatch, S.L., Keller, S.L.: {Separation of liquid phases in giant vesicles of
  ternary mixtures of phospholipids and cholesterol}.
\newblock {Biophys. J.} \textbf{{85}}({5}), {3074--3083} ({2003})

\bibitem{VK05}
Veatch, S.L., Keller, S.L.: Seeing spots: Complex phase behavior in simple
  membranes.
\newblock Biochimica et Biophysica Acta \textbf{1746}, 172--185 (2005)

\bibitem{VSK07}
Veatch, S.L., Soubias, O., Keller, S.L., Gawrisch, K.: Critical fluctuations in
  domain-forming lipid mixtures.
\newblock PNAS \textbf{104}, 17,650--17,655 (2007)

\bibitem{VSS05}
Venturoli, M., Smit, B., Sperotto, M.M.: Simulation studies of protein-induced
  bilayer deformations, and lipid-induced protein tilting, on a mesoscopic
  model for lipid bilayers with embedded proteins.
\newblock Biophys. J. \textbf{88}, 1778 (2005)

\bibitem{VenturoliSperottoKranenburgSmit06}
Venturoli, M., Sperotto, M.M., Kranenburg, M., Smit, B.: {Mesoscopic models of
  biological membranes}.
\newblock {Phys. Rep.} \textbf{{437}}({1-2}), {1--54} ({2006})

\bibitem{VSM03}
Vereb, G., Sz\"ollosi, J., Matko, J., Nagy, P., Farkas, T., Vigh, L., Matyus,
  L., Waldmann, T.A., Damjanovich, S.: Dynamic, yet structured: The cell
  membranes three decades after the singer-nicolson model.
\newblock PNAS \textbf{100}, 8053--8058 (2003)

\bibitem{VILL10.1}
Villa, A., Peter, C., van~der Vegt, N.F.A.: {Transferability of Nonbonded
  Interaction Potentials for Coarse-Grained Simulations: Benzene in Water}.
\newblock J. Chem. Theory Comput. \textbf{6}, 2434--2444 (2010)

\bibitem{VGD08}
Vostrikov, V.V., Grant, C.V., Daily, A.E., Opella, S.J., Koeppe~II, R.E.:
  Comparison of ''polarization inversion with spin exchange at magic angle''
  and ''geometric analysis of labeled alanines'' methods for transmembrane
  helix alignment.
\newblock J. Am. Chem. Soc. \textbf{130}, 12,584--12,585 (2008)

\bibitem{VK11}
Vostrikov, V.V., Koeppe~II, R.E.: Response of {GWALP} transmembrane peptides to
  changes in the tryptophan anchor positions.
\newblock Biochemistry \textbf{50}, 7522--7535 (2011)

\bibitem{Voth-CG-book}
Voth, G.A. (ed.): Coarse-Graining of Condensed Phase and Biomolecular Systems,
  1 edn.
\newblock CRC Press, Boca Raton (2008)

\bibitem{VYM05}
de~Vries, A.H., Yefimov, S., Mark, A.E., Marrink, S.J.: Molecular structure of
  the lecithin ripple phase.
\newblock PNAS \textbf{102}, 5392--5396 (2005)

\bibitem{WangHuZhang}
Wang, H., Hu, D., Zhang, P.: Measuring the spontaneous curvature of bilayer
  membranes by molecular dynamics simulations.
\newblock Commun. Comput. Phys. \textbf{13}, 1093--1106 (2013)

\bibitem{Wang10b}
Wang, Z.J., Deserno, M.: Systematic implicit solvent coarse-graining of bilayer
  membranes: lipid and phase transferability of the force field.
\newblock {N}ew {J}. {P}hys. \textbf{12}, 095,004 (2010)

\bibitem{Wang10a}
Wang, Z.J., Deserno, M.: A systematically coarse-grained solvent-free model for
  quantitative phospholipid bilayer simulation.
\newblock {J}. {P}hys. {C}hem. B \textbf{114}(34), 11,207--11,220 (2010)

\bibitem{WF05}
Wang, Z.J., Frenkel, D.: Modeling flexible amphiphilic bilayers: A solvent-free
  off-lattice {M}onte {C}arlo study.
\newblock J. Chem. Phys. \textbf{122}, 234,711 (2005)

\bibitem{Watson12}
Watson, M.C., Brandt, E.G., Welch, P.M., Brown, F.L.H.: Determining biomembrane
  bending rigidities from simulations of modest size.
\newblock Phys. Rev. Lett. \textbf{109}, 028,102 (2012)

\bibitem{Watson10}
Watson, M.C., Brown, F.L.H.: Interpreting membrane scattering experiments at
  the mesoscale: the contribution of dissipation within the bilayer.
\newblock {B}iophys. {J}. \textbf{98}(6), L09--L11 (2010)

\bibitem{WPW13}
Watson, M.C., Morriss-Andrews, A., Welch, P.M., Brown, F.L.H.: Thermal
  fluctuations in shape, thickness, and molecular orientation in lipid bilayers
  {II}: Finite surface tensions.
\newblock preprint  (2013)

\bibitem{WPW11}
Watson, M.C., Penev, E.S., Welch, P.M., Brown, F.L.H.: Thermal fluctuations in
  shape, thickness, and molecular orientation in lipid bilayers.
\newblock J. Chem. Phys. \textbf{135}, 244,701 (2011)

\bibitem{WeiklKozlovHelfrich98}
Weikl, T.R., Kozlov, M.M., Helfrich, W.: Interaction of conical membrane
  inclusions: Effect of lateral tension.
\newblock Phys. Rev. E \textbf{57}, 6988--6995 (1998).
\newblock \doi{10.1103/PhysRevE.57.6988}

\bibitem{WBS09}
West, B., Brown, F.L.H., Schmid, F.: Membrane-protein interactions in a generic
  coarse-grained model for lipid bilayers.
\newblock Biophys. J. \textbf{96}, 101--115 (2009)

\bibitem{WestSchmid10}
West, B., Schmid, F.: {Fluctuations and elastic properties of lipid membranes
  in the fluid and gel state: A coarse-grained Monte Carlo study}.
\newblock Soft Matter \textbf{6}, 1275--1280 (2010)

\bibitem{Wohlert06}
Wohlert, J., den Otter, W.K., Edholm, O., Briels, W.J.: Free energy of a
  trans-membrane pore calculated from atomistic molecular dynamics simulations.
\newblock {J}. {C}hem. {P}hys. \textbf{124}(15), 154,905 (2006)

\bibitem{WuMcConnel75}
Wu, S.H.W., McConnell, H.M.: Phase separations in phospholipid membranes.
\newblock Biochem. \textbf{14}(4), 847--854 (1975)

\bibitem{YBS10}
Yamamoto, T., Brewster, R., Safran, S.A.: Chain ordering of hybrid lipids can
  stabilize domains in saturated/hybrid/cholesterol lipid membranes.
\newblock EPL \textbf{91}, 28,002 (2010)

\bibitem{Yang.2003}
Yang, C., Horn, R., Paulsen, H.: The light-harvesting chlorophyll a/b complex
  can be reconstituted in vitro from its completely unfolded apoprotein.
\newblock Biochemistry \textbf{42}, 4527--4533 (2003)

\bibitem{YW07}
Yethiraj, A., Weisshaar, J.C.: Why are lipid rafts not observed in vivo?
\newblock Biophys. J. \textbf{93}, 3113--3119 (2007)

\bibitem{Yildiz.2012}
Yildiz, A.A., Knoll, W., Gennis, R.B., Sinner, E.K.: Cell-free synthesis of
  cytochrome bo(3) ubiquinol oxidase in artificial membranes.
\newblock Analytical Biochemistry \textbf{423}(1), 39--45 (2012).
\newblock \doi{10.1016/j.ab.2012.01.007}

\bibitem{YolDes-membrane}
Yolcu, C., Deserno, M.: {Membrane-mediated interactions between rigid
  inclusions: An effective field theory}.
\newblock Phys. Rev. E \textbf{86}, 031,906 (2012)

\bibitem{YolRotDesEPL}
Yolcu, C., Rothstein, I.Z., Deserno, M.: Effective field theory approach to
  {C}asimir interactions on soft matter surfaces.
\newblock Europhys. Lett. \textbf{96}, 20,003 (2011)

\bibitem{YolRotDes-longfilm}
Yolcu, C., Rothstein, I.Z., Deserno, M.: {Effective field theory approach to
  fluctuation-induced forces between colloids at an interface}.
\newblock Phys. Rev. E \textbf{85}, 011,140 (2012)

\bibitem{Zhelev93}
Zhelev, D.V., Needham, D.: Tension-stabilized pores in giant
  vesicles---determination of pore-size and pore line tension.
\newblock {B}iochim. et {B}iophys. {A}cta \textbf{1147}(1), 89--104 (1993)

\bibitem{ZilmanGranek96}
Zilman, A.G., Granek, R.: Undulations and dynamic structure factor of
  membranes.
\newblock Phys. Rev. Lett. \textbf{77}, 4788--4791 (1996)

\end{thebibliography}

\end{document}